# Flow and transport characteristics of hydrogen in a square-spiral Knudsen pump with coupled thermal transpiration and Poiseuille flow

Yawei Huang [a, b], Jianjun Ye [a, b, *], Jian Xiao [a], Junlong Xie [a, b, *], Dinggen Li [a, b], Xuesong He [b]

[a] School of Energy and Power Engineering, Huazhong University of Science and Technology, Wuhan, 430074, China

[b] Hanjiang National Laboratory, Wuhan 430064, China

* **Corresponding author: Dr.** Jianjun Ye; **Dr.** Junlong Xie

**E-mail address:** hustyjj@hust.edu.cn (Jianjun Ye); jlxie@hust.edu.cn (Junlong Xie)

## Abstract

Gas transport driven by thermal transpiration under temperature gradients, known as a Knudsen pump, has shown strong potential in gas separation and delivery applications. However, its performance and operational stability are strongly affected by structural design, system integration, and manufacturing complexity, highlighting the need for efficient and manufacturable multi-stage configurations. In this study, a novel spiral-structured Knudsen pump is proposed. Within the microchannels, a temperature gradient drives gas from the low- to high-temperature region, inducing pressure rise and thermal transpiration. In the main channel, a pressure gradient drives gas from the high- to low-temperature region, forming Poiseuille flow. Numerical methods are employed to investigate the flow mechanisms, and the effects of varying the numbers of series and parallel microchannels on gas transmission behavior are evaluated. The results indicate that bends of the main channel lead to nonlinear variations in pressure and temperature, as well as flow deceleration in the curved regions. Increasing the number of serial channels reduces the pressure load across the microchannel, thereby enhancing the overall transport performance. For instance, a ten-stage serial configuration improves performance by 3.35% compared with the four-stage case, reaching $3.10 \times 10^{-7}$ kg·m$^{-1}$·s$^{-1}$. In contrast, increasing the number of parallel microchannels reduces pressure loading while enhancing gas rarefaction, thereby strengthening thermal transpiration. A five-stage parallel configuration yields an 70.80% improvement over the two-stage case, reaching $3.07 \times 10^{-7}$ kg·m$^{-1}$·s$^{-1}$. These results provide valuable guidance for the design and fabrication of high-performance Knudsen pumps.

## 1. Introduction

In the fields of gas separation, gas mass spectrometry, and gas transmission within Micro-Electro-Mechanical Systems (MEMS), the efficiency and accuracy of gas transmission are crucial for enabling functions such as detection, adjustment, and control of these devices [1,2]. The Knudsen pump, as a miniature device that utilizes thermal transpiration to drive gas transmission, can recover and utilize low-grade energy sources such as solar energy, geothermal energy, and industrial waste heat. Moreover, due to the absence of mechanical moving parts, it offers advantages such as high reliability and long lifespan [3,4]. However, the transmission performance and accuracy of Knudsen pumps are influenced by factors such as its structural parameters, serial stages, and parallel stages. Therefore, studying the structural design of Knudsen pumps and the impact of serial and parallel stages on flow behaviors is a key prerequisite for achieving high performance and precise transmission.

To study the flow and transmission behavior of Knudsen pumps with different designs, some scholars have used micro-nano fabrication techniques to create single-stage and multi-stage Knudsen pump systems. For example, Qin et al. fabricated a single-stage Knudsen pump with a volume of 4 $cm^3$ using a three-mask processing technique. The system successfully achieved the separation of an alkane mixture within 60 seconds [5]. Based on this study, they used processes such as Deep Reactive Ion Etching (DRIE), Atomic Layer Deposition (ALD), and Low Pressure Chemical Vapor Deposition (LPCVD) to manufacture a 4-stage stacked Knudsen pump, achieving a blockage pressure of 0.255 kPa [6]. Byambadorj et al. designed a multi-stage Knudsen pump with rectangular microchannels using a five-mask photolithography process, achieving transmission rates of approximately 0.36 sccm and 0.75 sccm [7,8]. Subsequently, they proposed a single-stage Knudsen pump with rectangular channels for use in a micro gas chromatograph, achieving a transmission rate of 0.041 sccm [9]. In recent years, the team has integrated three unidirectional Knudsen pumps, a pre-concentrator, a separation column, and a detector on a 15 × 15 $mm^2$ chip, enabling gas sampling and analysis. Analysis shows that the error in quantitative analysis results for the content in the mixture is within 8.5% [10]. Furthermore, An et al. successfully built a 162-stage Knudsen pump with serpentine flow channels using a five-mask single-wafer fabrication process. Experimental results showed that the system could achieve a compression ratio of 844 at an input power of 0.39 W [11]. Cheng et al. used 3D printing technology and mixed cellulose ester (MCE) membranes to fabricate a bidirectional flow Knudsen pump. The bidirectional flow Knudsen pump achieved a maximum flow rate of 0.39 sccm and a blocking pressure of 818.2 Pa with an input power of 2 W [12]. Meanwhile, some scholars have utilized porous media materials to fabricate Knudsen pumps, exploring the effects of pore size and stage number on transmission performance. For instance, Kosyanchuk et al. used cellulose ester porous membranes and 3D printing technology to fabricate single-stage and multi-stage Knudsen pumps, studying the influence of stage number and pore size on pump performance. The results indicated that as the stage number increased, the blockage pressure of the Knudsen pump increased linearly [13]. Xiao et al. fabricated a Knudsen pump using porous ceramics for operation in low-temperature environments and analyzed the pressurization performance of helium, nitrogen, and argon. The results showed that helium exhibited the best pressurization performance [14]. Kugimoto et al. combined thermal transpiration flow with the phase change of water to create a stationary heat pump system. The system achieved a maximum pressure difference of 57.6 Pa and a mass flow rate of 0.0484 mg/(s·cm²), expanding the application range of Knudsen pumps [15].

On the other hand, other scholars have explored the effects of structures such as serrated edges, bends, and non-uniform channels on the flow and transmission behaviors of Knudsen pumps from a numerical perspective. For example, Huang et al. employed a numerical method based on the Navier–Stokes–Fourier (NSF) equations with slip boundary corrections to investigate the flow characteristics of a multi-stage Knudsen pump under blockage conditions. The results indicate that the transport performance exhibits a nonlinear decreasing trend with increasing number of blockages [16]. Additionally, Wang et al. designed single-stage and multi-stage Knudsen pumps with triangular and rectangular cantilever beams and analyzed their operational performance using DSMC. The study showed that, compared to the original square cantilever beam, the Knudsen force with the triangular cantilever beam increased by 42.9% [17,18]. Sugimoto et al. used the DSMC method to perform a numerical analysis of a single-stage Knudsen pump with a three-dimensional rectangular microchannel structure. They explored the impact of boundary effects on the flow behaviors [19]. Furthermore, Bond et al. studied the flow and transmission behaviors of a sine-wave structured Knudsen pump based on the Unified Gas Kinetic Scheme (UGKS). The results showed that, compared to the rectangular structure, the sine-wave structure improved transmission performance [20]. The team of Bond also used UGKS to analyze flow behaviors in curved channels, finding that the Prandtl number had a significant impact on the transmission performance of argon and nitrogen [21]. Tatsios et al. analyzed the performance of a converging-diverging structure Knudsen pump based on a linearized dynamic model equation. They found that an increase in the channel inclination reduced the transmission performance of the Knudsen pump [22].

The above studies indicate that the structure of Knudsen pumps, as well as the number of serial and parallel stages, significantly affects the flow behaviors and transmission performance, providing important experimental data and theoretical support for optimizing designs. However, the structural design, level of integration, and manufacturing complexity of Knudsen pumps directly influence their application potential and operational stability. Therefore, developing an efficient, easily manufacturable, and easily integrable Knudsen pump is crucial for advancing its practical application and manufacturing.

This paper presents a novel multi-stage Knudsen pump with a spiral-shaped configuration, in which the internal flow is driven by the coupling of thermal transpiration and Poiseuille flows. A numerical model for the multi-stage Knudsen pump is developed based on the NSF equations, incorporating corrections for velocity slip and temperature jump. The flow mechanism of the spiral-shaped Knudsen pump is investigated. Furthermore, the effects of the number of serial and parallel microchannels on the flow and transmission behavior of Knudsen pumps are analyzed, providing valuable model references for the practical application and manufacturing of Knudsen pumps.

## 2. Numerical method for the spiral-shaped Knudsen pump

### 2.1 Geometric model of Knudsen pump

The schematic diagram and dimensions of the rectangular channel with spiral-shaped configuration are shown in Fig. 1. It mainly consists of microchannels and main channels. The microchannels are located at the central position of the main channel, labeled from Location 1 to Location 10. The gray color represents solid walls, and the blue color represents the hydrogen gas flow region. The structural parameters of the main channel are as follows: height $H$ = 10 μm and length $L$ = 780 μm. Due to the bends in the main channel, the container length is further divided into 13 sections, labeled $L1$ to $L13$, as detailed in Table 1. The wall thickness is $\delta$ = 1μm. The microchannel parameters are: length $l$ = 25 μm and thickness $\varepsilon$ = 0.5 μm. The origin of the model is set at the center of the Knudsen pump inlet, denoted as point $O$. The $x$-axis is along the gas flow direction, and the $y$-axis is along the container height direction. For boundary conditions, the relative pressures at the inlet and outlet are both set to atmospheric pressure, i.e., $P_{in} = P_{out} = 0$ Pa. The reference pressure used in the computation is set to 1 standard atmosphere. The inlet solid wall of the microchannel is maintained as a low-temperature boundary at $T_0$ = 273 K, while the outlet solid wall is specified as a high-temperature boundary at $T_c$ = 323 K, corresponding to a temperature difference of ΔT=50 K. The wall temperature within the microchannels of the spiral-shaped Knudsen pump is determined via heat conduction analysis, thereby providing a more realistic representation of the actual operating conditions. The outer wall of the pump is assumed to be adiabatic. Numerical results indicate that the wall thickness has a negligible effect on the internal temperature distribution. The wall material is solid silicon, and the working medium is hydrogen gas. Under this configuration, the imposed temperature gradient induces a thermal transpiration flow in the microchannel from the cold side to the hot side, resulting in the formation of a high-pressure region at the microchannel outlet. However, the elevated pressure developed at the hot side of the microchannel drives a Poiseuille flow in the main channel from the hot side to the cold side.

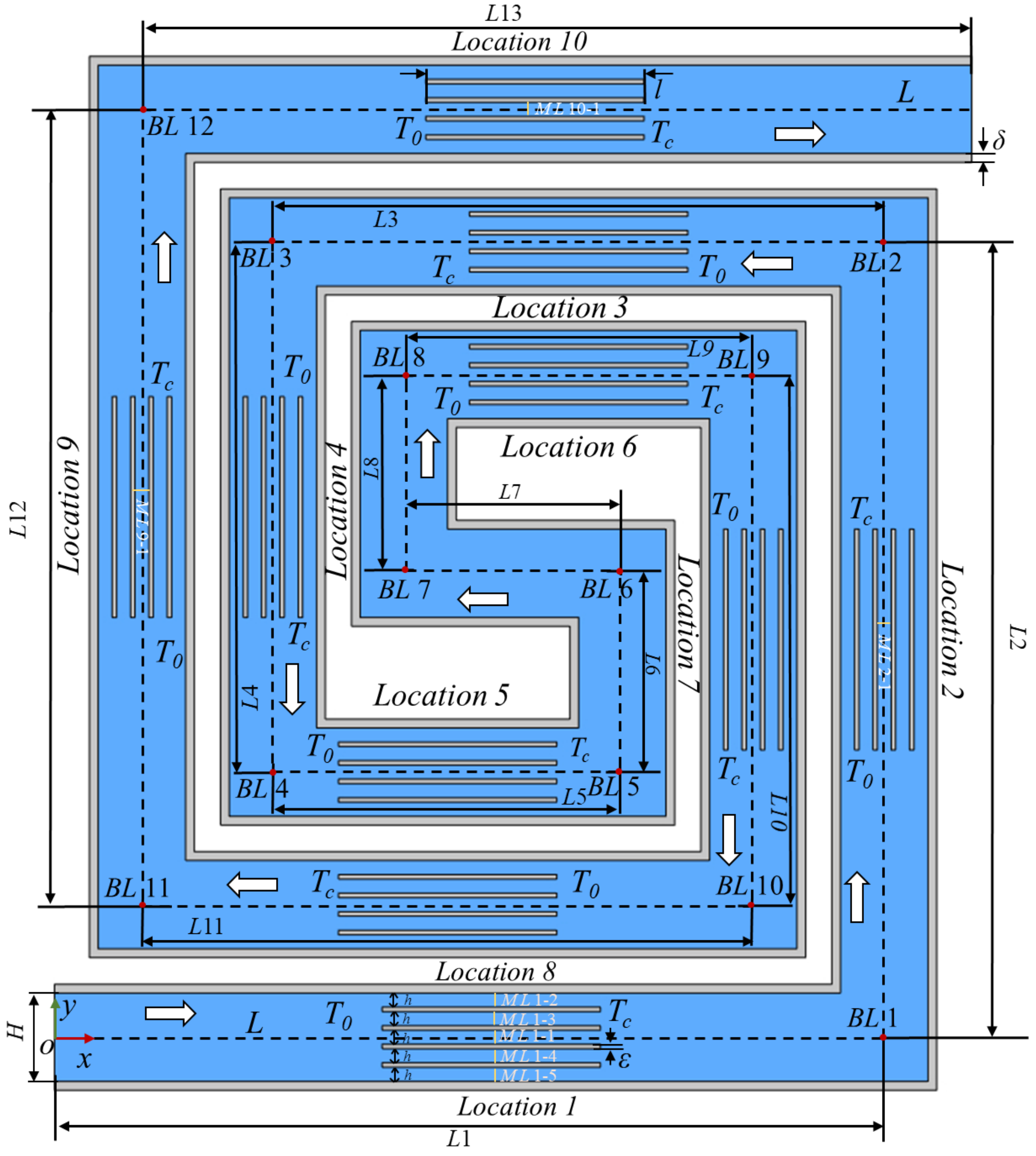

Fig. 1. Multi-stage Knudsen pump with a spiral-shaped structure.

To monitor the flow behavior within the hydrogen Knudsen pump, multiple monitoring lines and points are set inside the pump. The monitoring points are located at the bends of the Knudsen pump, labeled as Bend Location (*BL*), ranging from *BL* 1 to *BL* 12, and are used to monitor the changes in flow behaviors caused by the curvature of the flow channels. The monitoring lines (*ML*) are located within the microchannels. The position of the first-stage microchannel is at Location 1, with the monitoring lines arranged from top to bottom as *ML* 1-2, *ML* 1-3, *ML* 1-1, *ML* 1-4, and *ML* 1-5. A monitoring line is placed at the central microchannel position of Location 2, Location 9, and Location 10, labeled as *ML* 2-1, *ML* 9-1, and *ML* 10-1, respectively. These lines are used to compare the differences in flow behaviors between different stages.

Table 1 Dimensions of the Knudsen pump.

| Designation | Length (μm) | Designation | Length (μm) |
|---|---|---|---|
| *l* | 25 | *L*6 | 22.5 |
| H | 10 | *L*7 | 25 |
| δ | 1 | *L*8 | 22.5 |
| ε | 0.5 | *L*9 | 40 |
| *L*1 | 95 | *L*10 | 60 |
| *L*2 | 90 | *L*11 | 70 |
| *L*3 | 70 | *L*12 | 90 |
| *L*4 | 60 | L13 | 95 |
| *L*5 | 40 | Total Length (L) | 780 |

To study the flow mechanism of the spiral-shaped Knudsen pump and the effects of serial stages (Number of Series Stages, NSS) and parallel stages (Number of Parallel Stages, NPS) on the flow and transmission behaviors within the pump, Table 2 presents the computational conditions for different research cases. Case 1 is used to analyze the flow mechanisms of the Knudsen pump, with NPS and NSS set to 5 and 4, respectively. The microchannel height is $h$ = 1.6 μm, and the microchannel positions are at Locations 1, 2, 9, and 10. Case 2, Case 3, and Case 4 are used to analyze the effects of serial stages on the flow and transmission behaviors. NPS is set to 5, with the microchannel height $h$ = 1.6 μm. NSS is set to 6, 8, and 10, respectively. In Case 2, the microchannel positions are at Locations 1, 2, 5, 6, 9, and 10. In Case 3, the microchannel positions are Locations 1, 2, 4, 5, 6, 7, 9, and 10. In Case 4, the microchannel positions are Locations 1 to 10. Case 5, Case 6, Case 7, and Case 8 are used to analyze the effects of parallel stages on the flow and transmission behaviors. NSS is set to 6. NPS is set to 2, 3, 4, and 5, respectively, with corresponding microchannel heights $h$ of 4.75 μm, 3 μm, 2.125 μm, and 1.6 μm. The microchannel positions are at Locations 1, 2, 4, 7, 9, and 10.

The effects of the number of parallel microchannels in Cases 5–8 are evaluated while maintaining a fixed characteristic dimension of the main channel, thereby elucidating the resulting variations in the flow and transmission characteristics of the Knudsen pump. To provide a complementary assessment, the influence of the number of parallel microchannels is further investigated while maintaining a constant characteristic dimension of the individual microchannels, as illustrated in Fig. S1 of the Supporting Information. Accordingly, Comparison Cases 1–3 are introduced to systematically evaluate the effects of varying the number of parallel microchannels on the flow and transmission characteristics under an identical microchannel characteristic dimension. For all comparison cases, the height of each microchannel is fixed at 1.6 μm, thereby maintaining the same degree of gas rarefaction. Comparison Cases 1, 2, and 3 correspond to configurations comprising 2, 3, and 4 parallel microchannels, respectively. The microchannels in these configurations are arranged at selected positions among Locations 1, 2, 4, 7, 9, and 10 according to the prescribed geometric configurations, as detailed in Table S1.

Table 2 Computational cases for the spiral-shaped Knudsen pump.

| Case | Research issue | Number of Parallel Stages (NPS) | Microchannel Characteristic Dimension $h$ (μm) | Number of Serial Stages (NSS) | Locations |
|---|---|---|---|---|---|
| Case 1 | Flow mechanism | 5 | 1.6 | 4 | Locations 1, 2, 9, and 10 |
| Case 2 | Effects of serial stages on flow and transmission behaviors | | | 6 | Locations 1, 2, 5, 6, 9, and 10 |
| Case 3 | | | | 8 | Locations 1, 2, 4, 5, 6, 7, 9, and 10 |
| Case 4 | | | | 10 | Locations 1, 2, 3, 4, 5, 6, 7, 8, 9, and 10 |
| Case 5 | Effects of parallel stages on flow and transmission behaviors | 2 | 4.75 | 6 | Locations 1, 2, 4, 7, 9, and 10 |
| Case 6 | | 3 | 3 | | Locations 1, 2, 4, 7, 9, and 10 |
| Case 7 | | 4 | 2.125 | | Locations 1, 2, 4, 7, 9, and 10 |
| Case 8 | | 5 | 1.6 | | Locations 1, 2, 4, 7, 9, and 10 |

## 2.2 Control equations for the computational domain

In traditional continuum fluid dynamics simulations, the NSF equations are typically used to characterize the spatial distribution of macroscopic quantities such as velocity, temperature, and pressure, thereby analyzing and predicting the fluid flow behavior [23,24]. Under two-dimensional conditions, the expressions of the NSF equations are given by Eqs (1) – (6).

The Knudsen pump numerical model constructed in this study operates in the slip flow regime with a Knudsen number ($Kn$) ranging from 0.001 to 0.1 [25]. In this flow regime, although the NSF equations can still be used for solving [26], it is essential to consider the velocity slip and temperature jump phenomena at the boundaries caused by the rarefied gas effects. This helps in more accurately describing the flow behaviors of rarefied gases near the boundaries [27].

Continuity equation:

$$\nabla \cdot (\rho \boldsymbol{u}) = 0 \tag{1}$$

Momentum equation:

$$\rho(\boldsymbol{u} \cdot \nabla)\boldsymbol{u} = \nabla \cdot (-p\boldsymbol{I} + \tau) + \boldsymbol{F} \tag{2}$$

$$\tau = \mu(\nabla \boldsymbol{u} + (\nabla \boldsymbol{u})^T) - \frac{2}{3}\mu(\nabla \cdot \boldsymbol{u})\boldsymbol{I} \tag{3}$$

Energy equation:

$$\nabla \cdot \left[\rho \boldsymbol{u}\left(e + \frac{1}{2}\boldsymbol{u}^2\right)\right] = \nabla \cdot (k\nabla T) + \nabla \cdot (-p\boldsymbol{u} + \tau \cdot \boldsymbol{u}) + \boldsymbol{u} \cdot \boldsymbol{F} \tag{4}$$

Where, $\boldsymbol{u}$ is the velocity field of the fluid (m/s), which can be divided into two components: $u$ and $v$. $p$ is the pressure (Pa), and $T$ is the temperature (K). $\rho$ is the density of hydrogen gas (kg/m³). Under the assumption of ideal gas behavior, the pressure $p$ can be expressed in terms of density through the equation of state $p = \rho RT$, where R is the specific gas constant. $\boldsymbol{F}$ is the body force term (N/m³), which refers to the gravitational force in the $y$-direction in this study. $\mu$ is the dynamic viscosity of hydrogen gas (Pa·s), reflecting the viscous resistance to the internal flow of the gas. $e$ is the internal energy per unit mass of the fluid (J/kg), can be expressed as a function of the specific heat capacity at constant volume, $C_V$ (J/(kg·K)), and the temperature $T$, i.e., $e = C_V T$. $\frac{1}{2}\boldsymbol{u}^2$ is the kinetic energy per unit mass of the fluid (J/kg). The total energy per unit mass of the fluid is given by $e+\frac{1}{2}\boldsymbol{u}^2$. $k$ is the thermal conductivity (W/(m·K)), and $\boldsymbol{I}$ is the unit tensor. $\tau$ is the viscous stress tensor (Pa), which can be expressed as:

$$\tau = \begin{bmatrix} \tau_{xx} & \tau_{xy} \\ \tau_{yx} & \tau_{yy} \end{bmatrix} \tag{5}$$

The viscous stress tensor $\tau$ can be decomposed into the following four equations:

$$\tau_{xx} = 2\mu\frac{\partial u}{\partial x}, \qquad \tau_{yy} = 2\mu\frac{\partial v}{\partial y}, \qquad \tau_{xy} = \tau_{yx} = \mu(\frac{\partial u}{\partial y}+\frac{\partial v}{\partial x}) \tag{6}$$

The temperature gradient within the channel causes the gas to flow from the low-temperature end to the high-temperature end, which is known as the thermal transpiration effect. To account for the influence of the thermal transpiration effect on gas flow [28], we introduce slip boundary conditions with thermal transpiration correction, as shown in Eqs. (7) – (9).

$$u_s = \frac{2-\sigma_v}{\sigma_v}\frac{\lambda}{\mu}\tau_{n,t} + 0.75\frac{\mu}{\rho T_g}\nabla_t T_g \tag{7}$$

$$\tau_{n,t} = \tau\boldsymbol{n} - \left((\boldsymbol{n}^T\tau\boldsymbol{n})\boldsymbol{n}\right) \tag{8}$$

$$\nabla_t T_g = \nabla T_g - \left(\boldsymbol{n}\cdot\nabla T_g\right)\boldsymbol{n} \tag{9}$$

In equation (7), $u_s$ represents the slip velocity of the gas at the wall (m/s), and $\lambda$ is the mean free path of the gas (m). $\tau_{n,t}$ is the tangential component of the stress tensor near the wall. $\nabla_t T_g$ is the tangential component of the gas temperature gradient. The tangential momentum adjustment coefficient $\sigma_v$ is used to describe the degree of momentum transfer between the gas molecules and the solid surface during collisions [29]. In this study, the value of $\sigma_v$ is taken as 0.9.

Meanwhile, the wall temperature jump equation can be expressed as shown in Eq (10) [30]:

$$T_j = \xi_T \lambda \boldsymbol{n} \cdot \nabla T_g \tag{10}$$

In Eq (10), $T_j$ represents the temperature jump value near the wall (K). $\boldsymbol{n} \cdot \nabla T_g$ is the normal component of the gas temperature gradient near the wall (K/m). $\xi_T$ is given by $\frac{2-\sigma_T}{\sigma_T}\frac{2\gamma}{\gamma+1}\frac{k}{\mu C_p}$, where $\gamma$ is the specific heat ratio of the gas, and $C_p$ is the specific heat capacity at constant pressure (J/(kg·K)). $\sigma_T$ is the thermal accommodation coefficient, which characterizes the degree of energy exchange between gas molecules and the solid surface during collisions. Experimental measurements by Velthuis et al. showed that the tangential momentum accommodation coefficient of hydrogen varies with the wall material, with reported values ranging from approximately 0.78 to 0.99 [31]. According to the Maxwell gas–surface interaction model discussed by Sharipov, the momentum and thermal accommodation coefficients can be represented by the same accommodation parameter[32]. This simplification is adopted in the present study. Based on the boundary-condition settings used in previous studies of hydrogen Knudsen compressors and low-temperature-driven multi-stage Knudsen pumps, $\sigma_v = \sigma_T$ is selected as a representative model parameter [25,33–35].

To assess the rarefied state of the gas in the microchannel, the Knudsen number can be calculated using the gas constant $R$ (J/(kg·K)), pressure $p$ (Pa), molecular molar mass $m$ (kg), dynamic viscosity $\mu$ (Pa·s), temperature $T$ (K), and the characteristic dimension of the channel $D$ (m), as shown in Eq (11) [36].

$$Kn = \frac{\lambda}{D} = \frac{\mu}{p}\sqrt{\frac{\pi RT}{2m}} \cdot \frac{1}{D} \tag{11}$$

The transmission performance of the Knudsen pump is characterized by the mass flow rate per unit depth, $M$ (kg·m$^{-1}$·s$^{-1}$), with an out-of-plane depth of 1 m. It is calculated by integrating the product of gas density ($\rho$) and velocity ($u$) over the flow cross section of each microchannel at Location 1 and summing the contributions from all parallel microchannels, where $N$ denotes the total number of parallel microchannels, as expressed in Eq. (12).

$$M = \sum_{i=1}^{N} \int_0^h \rho u dy \tag{12}$$

For comparison with existing experimental results, Eq. (13) gives the conversion between the mass flow rate and the standard volumetric flow rate. For practical devices, when the out-of-plane depth $B$ is much larger than the microchannel height, wall effects in the out-of-plane direction can be neglected. The total mass flow rate can then be expressed as $M_{tot} = BM$. The corresponding standard volumetric flow rate is obtained by dividing the total mass flow rate by the gas density $\rho_{std}$ at the specified standard conditions.

$$Q = 60 \times 10^6 \frac{M_{tot}}{\rho_{std}} = 6 \times 10^7 \frac{BM}{\rho_{std}} \tag{13}$$

Previous studies have typically normalized the standard volumetric flow rate by the total effective opening area of the microchannels [7]. For a Knudsen pump with $N$ parallel microchannels, each with a height $h$, the total effective opening area is given by $A_{eff} = NhB$. Accordingly, the standard volumetric flow rate per unit effective opening area $Q_{sccm}$ (sccm·mm$^{-2}$) can be further calculated using Eq. (14).

$$Q_{sccm} = \frac{Q}{A_{eff}} = 6 \times 10^7 \frac{BM/\rho_{std}}{10^6 NhB} = 60 \frac{M}{\rho_{std} Nh} \tag{14}$$

### 2.3 Validation of the Knudsen pump numerical model

In the rarefied gas flow module of COMSOL Multiphysics, the built-in slip flow formulation is adopted to characterize thermal transpiration induced by temperature gradients. The governing equations are discretized within a finite element framework. In this study, a fully coupled solution strategy is employed, in which the pressure and velocity fields are solved simultaneously as a unified variable system. Compared with segregated iterative schemes, the fully coupled approach exhibits improved numerical stability and convergence behavior, while also reducing the number of iterations and improving computational efficiency. For the resulting discretized linear algebraic system, the direct solver PARDISO, based on LU decomposition, is employed. This solver is among the most efficient direct algorithms integrated within COMSOL. To ensure numerical accuracy and robustness, the residual convergence criterion is set to $1\times10^{-6}$. The solution is deemed converged when the residuals of all dependent variables fall below this prescribed threshold.

Based on the numerical model of the spiral-shaped Knudsen pump and the computational cases, Case 1 is used as an example for grid generation and grid independence verification, as shown in Fig. 2. The model uses a structured hexahedral mesh, with mesh refinement near the wall surfaces, as shown in Fig. 2 (a). The transmission performance of the Knudsen pump is selected as the criterion for grid independence, as shown in Fig. 2 (b). When the total number of mesh elements exceeds $3\times10^5$, the mass flow rate stabilizes at $3.00 \times 10^{-7}$ kg·m$^{-1}$·s$^{-1}$. At this point, the numerical results become independent of the mesh size. Therefore, for subsequent computational cases, a mesh with approximately $4.5\times10^5$ elements is used for the calculations. Meanwhile, to further assess the sensitivity of the local pressure and velocity distributions to grid resolution, three grids containing approximately $1.6\times10^5$, $4.5\times10^5$, and $5.1\times10^5$ grid cells were selected to compare the streamwise pressure and velocity distributions along the *ML*1-1 microchannel, as shown in Figs. 2(c) and 2(d). The results indicate that the pressure and velocity profiles obtained using the grids with $4.5\times10^5$ and $5.1\times10^5$ grid cells are nearly identical, with relative differences in their mean values of only 0.03% and 0.01%, respectively. These results demonstrate that, once the number of grid cells reaches approximately $4.5\times10^5$, further grid refinement has a negligible effect on the mass flow rate and the pressure and velocity distributions within the channel, confirming the grid independence of the numerical results.

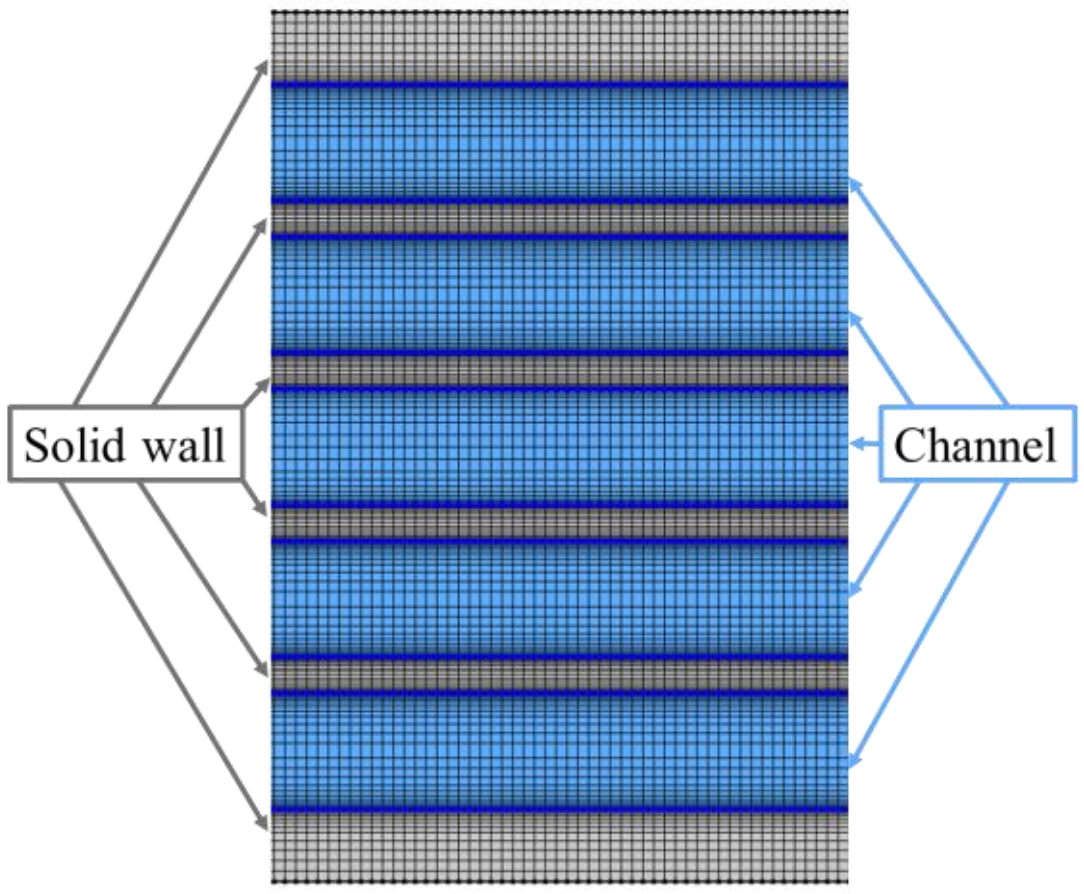

(a) Grid generation strategy for the Knudsen pump.

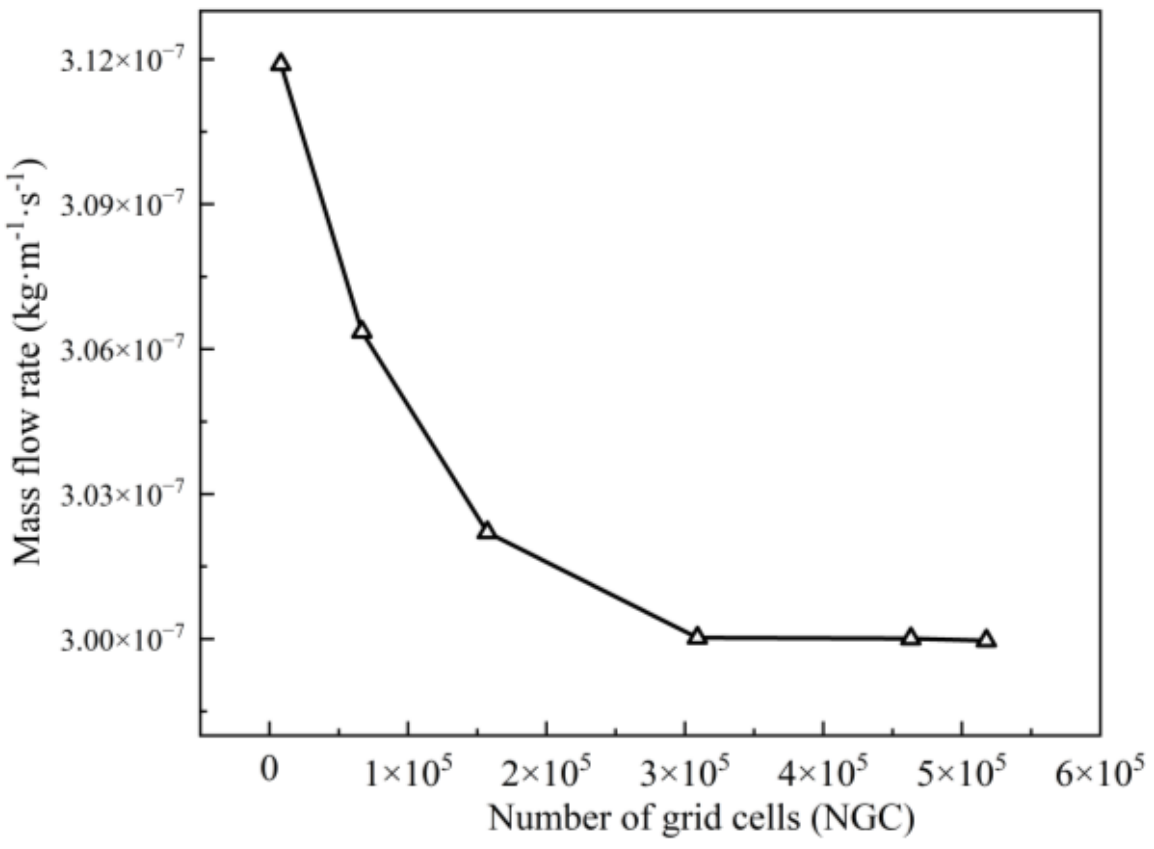

(b) Variation of mass flow rate with the number of grid cells.

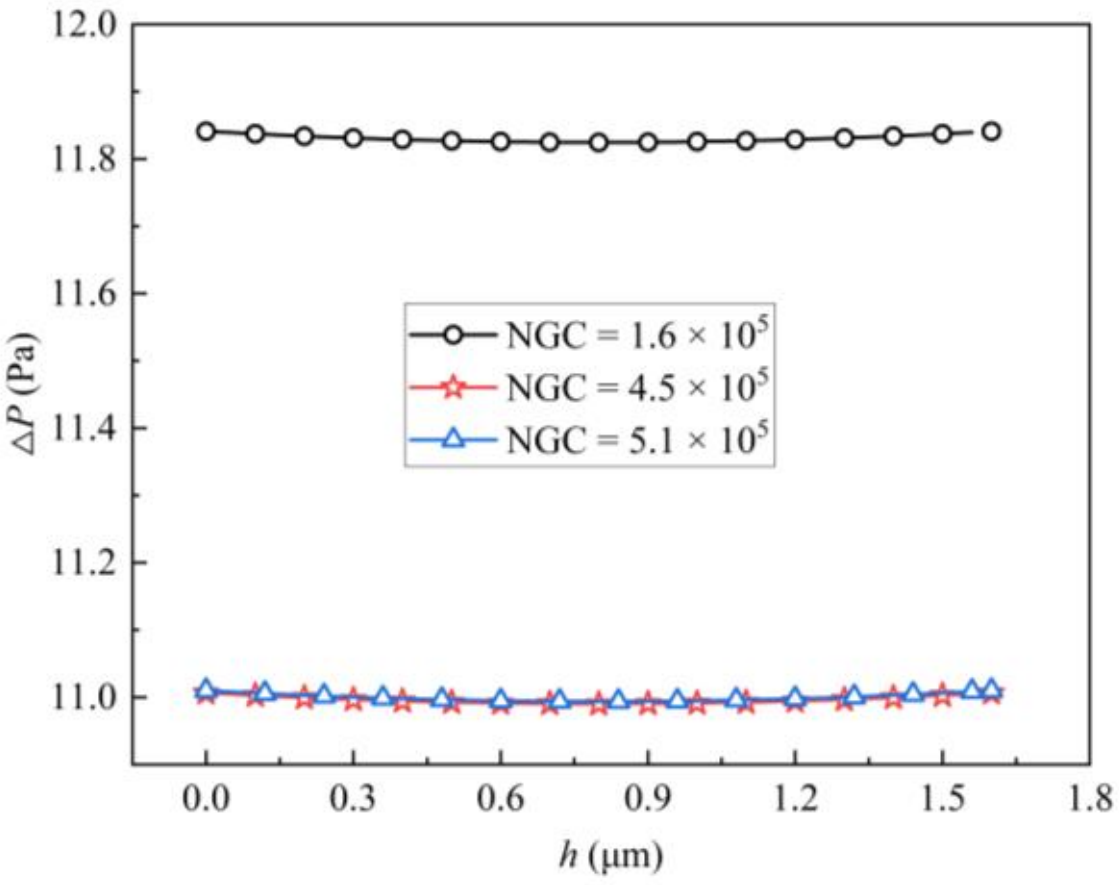

(c) Variation of pressure along ML1-1 with the number of grid cells.

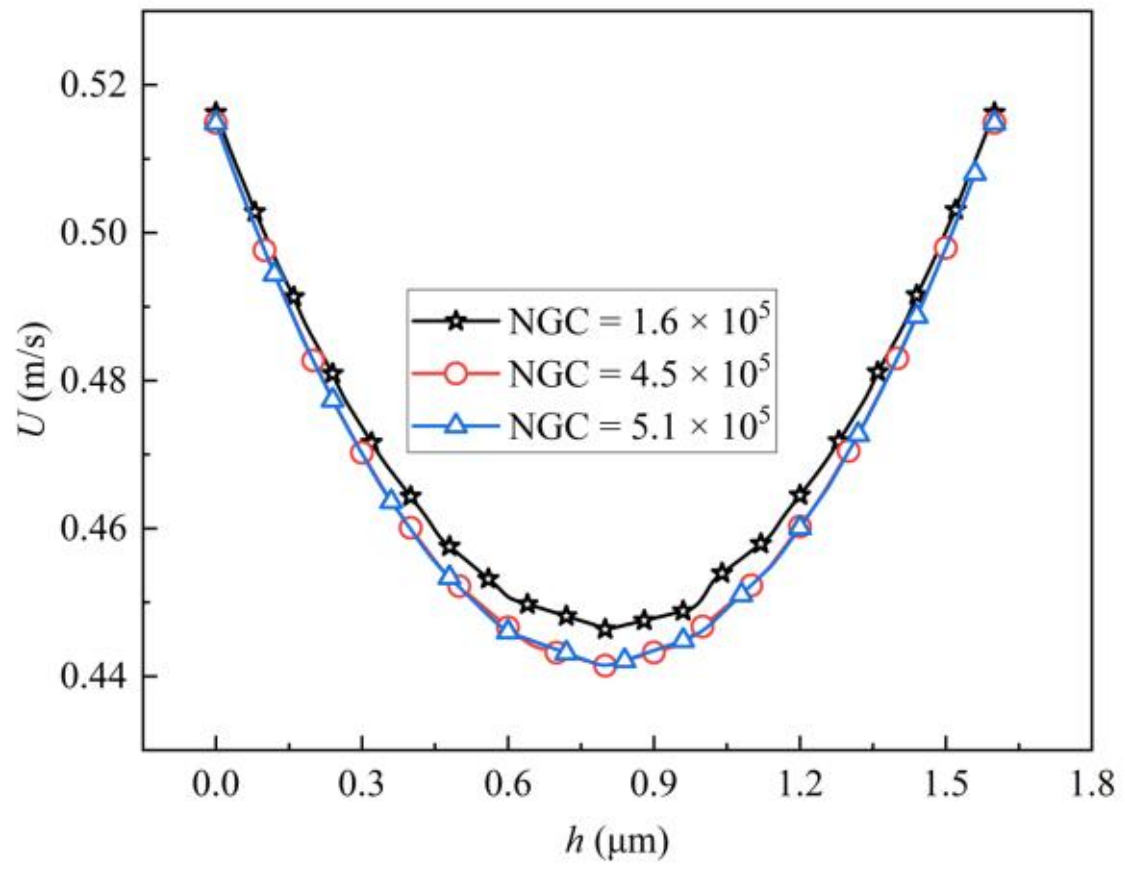

(d) Variation of velocity along ML1-1with the number of grid cells.

Fig. 2. Mesh generation and mesh independence validation for the numerical model of the Knudsen pump.

To validate the accuracy of the numerical method in predicting thermal transpiration and transmission performance in microchannels at different degrees of rarefaction within the slip-flow regime, the numerical results were compared with the experimental data reported by Yamaguchi et al. [37]. The experiment employed a rectangular microchannel with a height of 0.22 mm, a width of 6 mm, and a length of 73 mm. The two ends of the microchannel were connected to cold and hot reservoirs, respectively, to maintain stable thermal boundary conditions and enable pressure measurements. Argon was used as the working gas, with the hot and cold reservoirs maintained at 347.1 K and 289.2 K, respectively. Table 3 compares the experimental and numerical results at different mean rarefaction parameters ($\delta = \frac{pH}{\mu\sqrt{2RT}} \approx \frac{0.9025}{Kn}$). The results show that, with increasing mean rarefaction parameter, the deviations between the numerical and experimental results remain small. For example, when the Knudsen numbers are 0.067 and 0.089, the deviations between the experimental and numerical results are 0.27% and 0.71%, respectively. This confirms the reliability of the numerical method in predicting thermal transpiration and transmission behaviors in microchannels.

Table 3. Comparison of the present numerical results with the experimental data of Yamaguchi et al.

| | $\delta$ = 13.4 | $\delta$ = 10.1 | $\delta$ = 8.08 |
|---|---|---|---|
| Knudsen number | 0.067 | 0.089 | 0.11 |
| Experimental data of Yamaguchi et al. ($\times10^{-10}$ kg/s) | 0.746 | 0.700 | 0.651 |
| Numerical results obtained in this study ($\times10^{-10}$ kg/s) | 0.748 | 0.705 | 0.665 |
| Deviation (%) | 0.27 | 0.71 | 2.15 |

To verify the accuracy of the pressure-driven flow characteristics in curved channels, the numerical results were compared with both the rarefied gas flow results in the curved channel obtained by White et al. using the DSMC method [38] and the experimental results reported by Lee et al. [39]. The curved channels model established by White is shown in Fig. 3(a), with a channel length $L$ = 15 μm and a channel height $H$ = 1 μm. The inlet pressure is $P_{in}$ = 200000 Pa, and the inlet temperature is 300 K, while the outlet pressure is 66666 Pa. The comparison results are shown in Fig. 3(b). The gas pressure shows a nonlinear decrease along the channel, with a slight increase at the corner. On the other hand, the fluid velocity gradually increases, but there is a sudden drop in velocity at the corner. The average errors in pressure and Mach number are within 1% and 5%, respectively. The pressure and velocity trends are very similar to the results of White et al., which validates the effectiveness of the method used in this study.

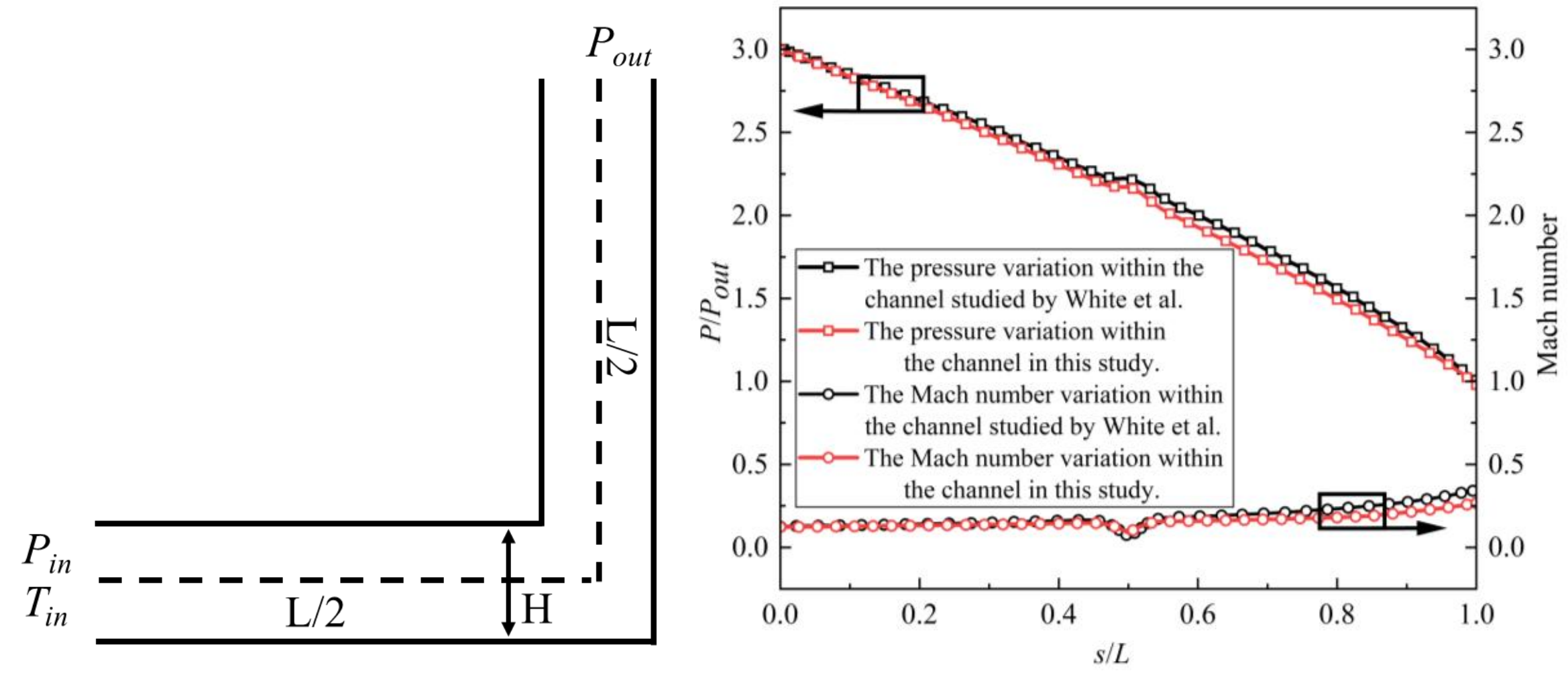

(a) The curved channel model proposed by White et al.

(b) Comparison with the results of the study by White et al.

Fig. 3. Comparison with the results of the curved channel study by White et al.

On the other hand, further validation was conducted by comparison with the experimental study of Lee et al., as shown in Fig. 4 [39]. Lee et al. constructed a double-bend channel model with dimensions of length × width × depth = 5810 μm × 20 μm × 1 μm. The bending angle is 45°, and the length of the bend is 50 μm, as shown in Fig. 4(a). In the experiment, the inlet and outlet boundary conditions were set as pressure boundaries, with an inlet pressure of 30.25 Psi and an outlet pressure of 0 Psi (atmospheric pressure). The working medium was argon. By using the model calculations in this study, the pressure distribution along the centerline of the channel was computed and compared with the experimental results, as shown in Fig. 4(b). The comparison shows that the numerical results align closely with the experimental data, with an average error of only 2%. This further validates the accuracy of the model presented in this study.

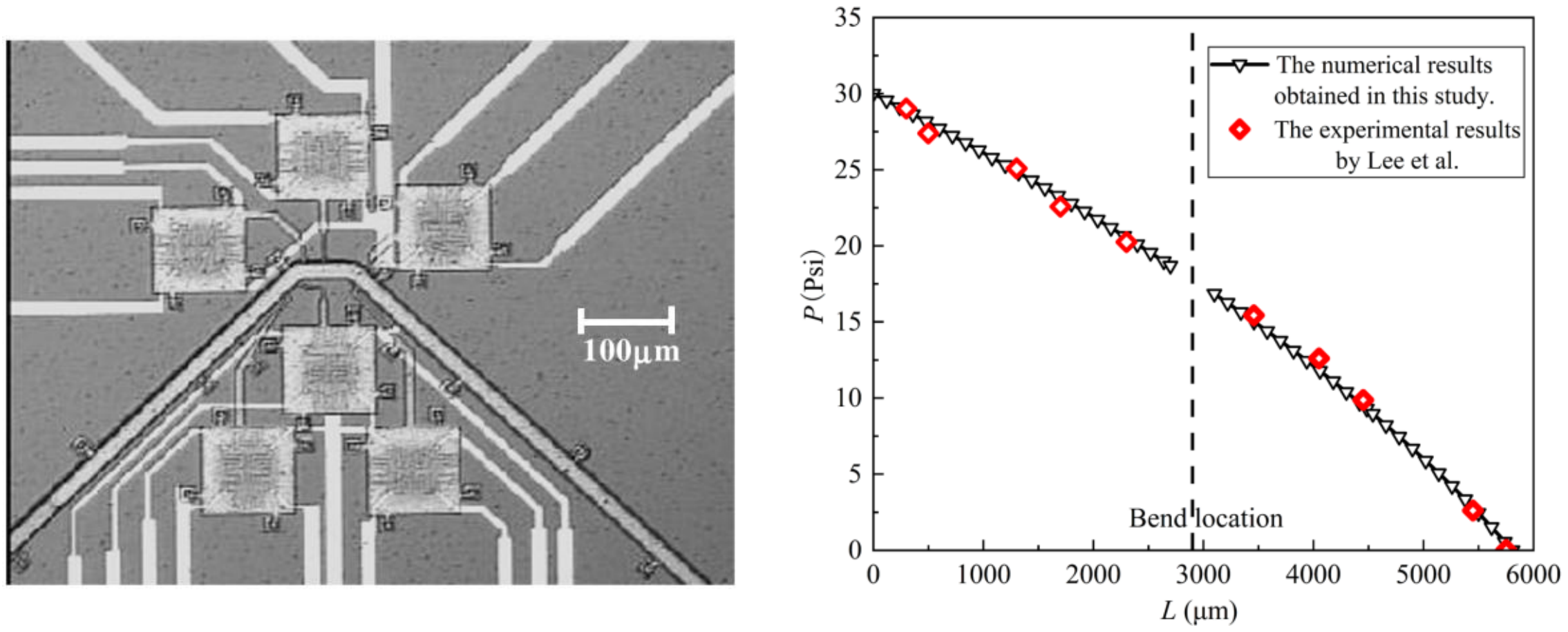

(a) The curved channel model studied by Lee et al.

(b) Comparison with the results of the study by Lee et al.

Fig. 4. Comparison with the results of the experimental study by Lee et al.

## 3. Results

### 3.1 Flow mechanism of the Knudsen pump

To investigate the flow mechanisms of the spiral-shaped Knudsen pump, Fig. 5 presents the temperature distribution and Knudsen number distribution within the pump. The temperature in the microchannels exhibits a periodic linear increase, while the temperature within the main channel gradually decreases. For instance, in the microchannels at Locations 1, 2, 9, and 10, the temperature rises from 273 K at the inlet to 323 K at the outlet. However, within the main channel, the temperature decreases from 323 K at the outlet of the microchannel to 273 K at the inlet of the next-stage microchannel. A nonlinear temperature change is observed near $s/L = 0.5$, which is primarily due to the frequent bends present in the middle of the Knudsen pump. These bends significantly alter the flow path. Flow curvature and acceleration increase local flow complexity, resulting in nonlinear temperature variations. On the other hand, the variation trend of the Knudsen number within the pump is similar to that of the temperature. In the microchannels at Locations 1, 2, 9, and 10, the Knudsen number increases from 0.064 at the inlet to 0.078 at the outlet. This indicates that the gas flow is in the slip flow regime, and the difference in the Knudsen number between the various stages of microchannels is small.

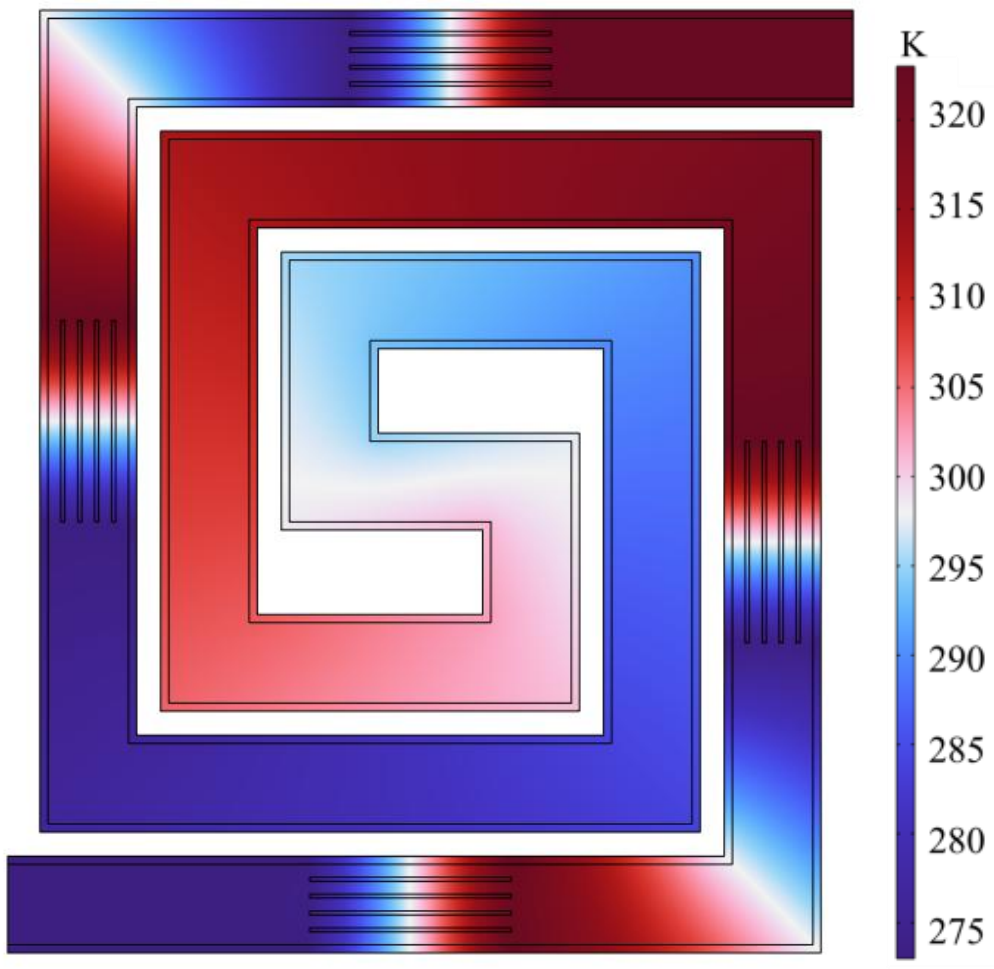

(a) Contour plot of temperature distribution.

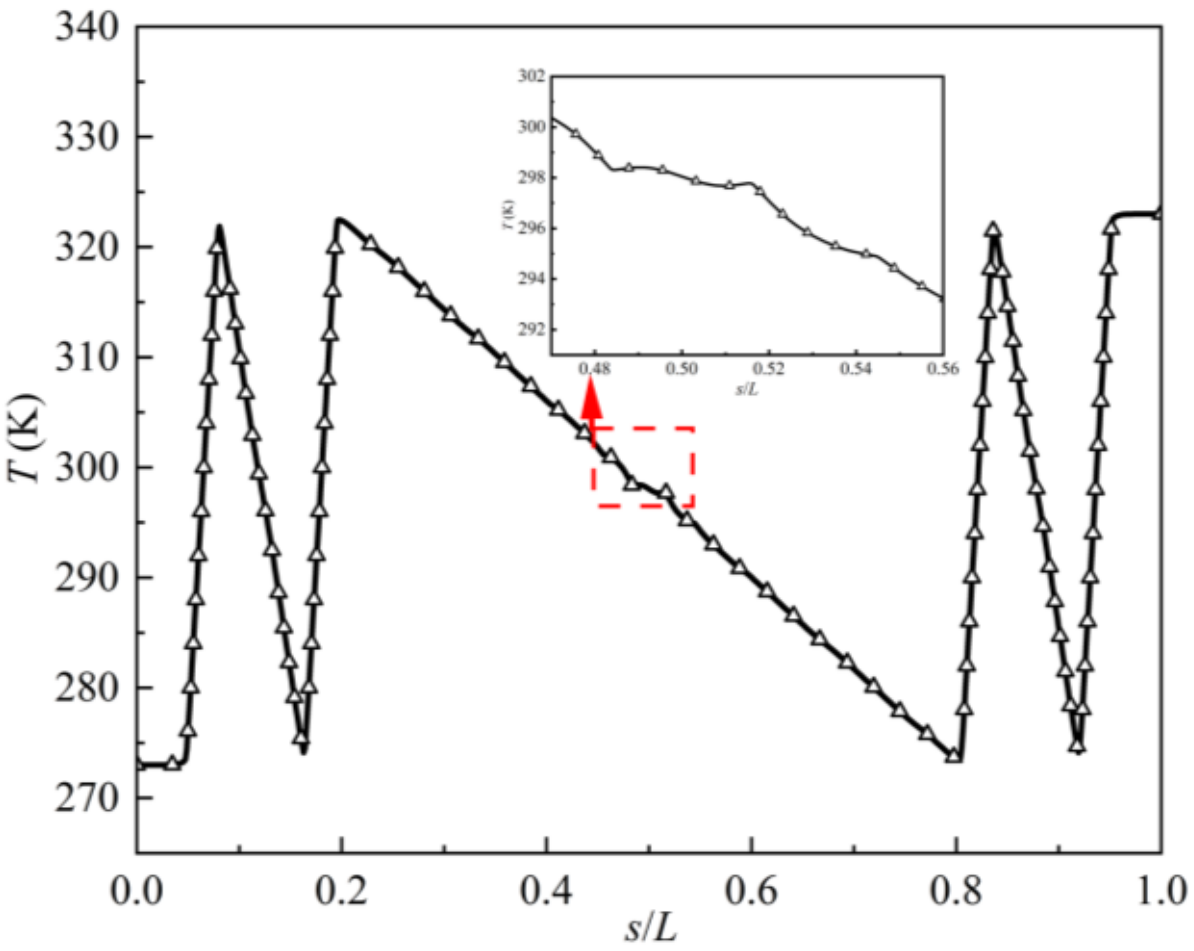

(b) Temperature distribution along the $L$-line.

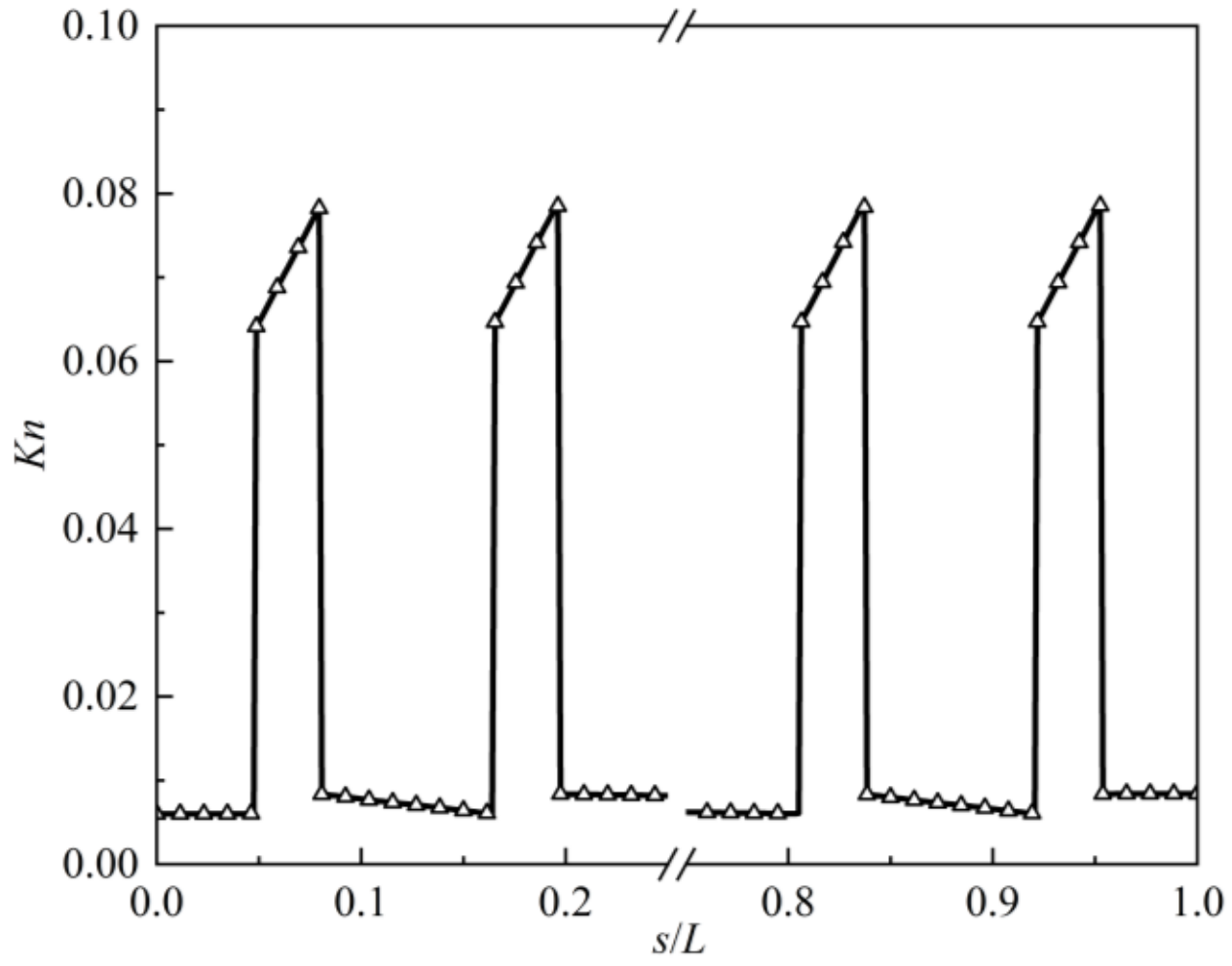

(c) Knudsen number distribution along the *L*-line.

Fig. 5. Temperature and Knudsen number distribution characteristics under Case 1 condition.

To study the flow behaviors of the Knudsen pump, Fig. 6 presents the pressure distribution within the pump. The results show that at Location 1 and Location 10, the pressure difference across the two ends of the microchannels exhibits an almost symmetric distribution. In the microchannels at Locations 2 and 9, there is a significant difference in the pressure difference at both ends. For instance, the pressure differences across the microchannels at Locations 1, 2, and 9 are 47 Pa, 86 Pa, and 77 Pa, respectively. Within the main channel, the pressure gradually decreases. However, at all bend locations within the main channel, a slight pressure increase is observed. For example, the pressure rises from 59.4 Pa to 59.7 Pa at *BL* 2. This is mainly attributed to the abrupt geometric change at the bend, which forces the gas flow to rapidly change direction and locally decelerate under wall confinement, resulting in a stagnation-like increase in static pressure. Overall, within the microchannels of the spiral-shaped Knudsen pump, the gas flows due to the thermal transpiration effect driven by the temperature gradient. In the main channel, the gas flow is driven by pressure.

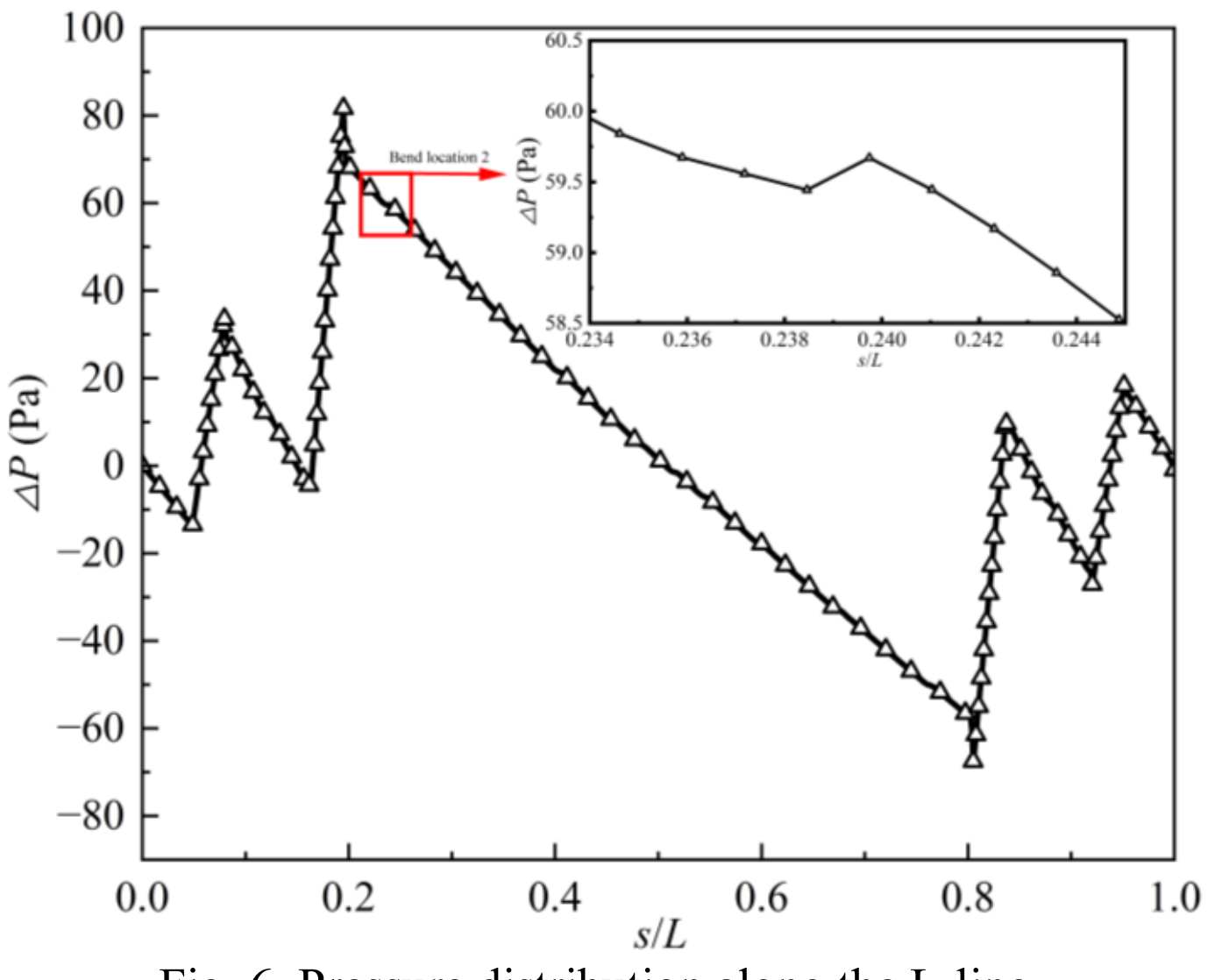

Fig. 6. Pressure distribution along the L-line.

Fig. 7 illustrates the velocity distribution within the Knudsen pump. The results show that the velocity within the microchannels increases along the flow direction. For example, in the microchannel at Location 1, the velocity increases from 0.41 m/s at the inlet to 0.48 m/s at the outlet, which corresponds to a 17% increase. In contrast, the velocity of the fluid within the main channel shows a decreasing trend. It is particularly notable that vortices form on the outer side of the main channel at all bends, which are referred to as Moffatt vortices. This is mainly because the local characteristic length scale near the corners is very small, rendering inertial effects negligible and causing the flow to be dominated by viscous effects. The outer main flow in the central region of the main channel transfers momentum into the corner regions through viscous shear. Meanwhile, the slip velocities along the two walls adjacent to the outer corner are very small. Under these constraints, the gas cannot flow directly through the corner region and instead forms a closed recirculating flow, resulting in Moffatt vortices [40]. After the fluid passes the bend, the velocity recovers to the level prior to the bend. On the other hand, the averaged flow velocity in the curved section of the main channel is generally lower than that in the straight section. For example, the averaged flow velocity in the curved region is 0.23 m/s, whereas it is approximately 0.40 m/s in the upstream straight section of the main channel. The sudden drop in velocity, as well as the reduction in the average flow velocity in the curved section, is primarily attributed to the increase in the effective flow cross-sectional area relative to the straight segment, which leads to a decrease in the local flow velocity.

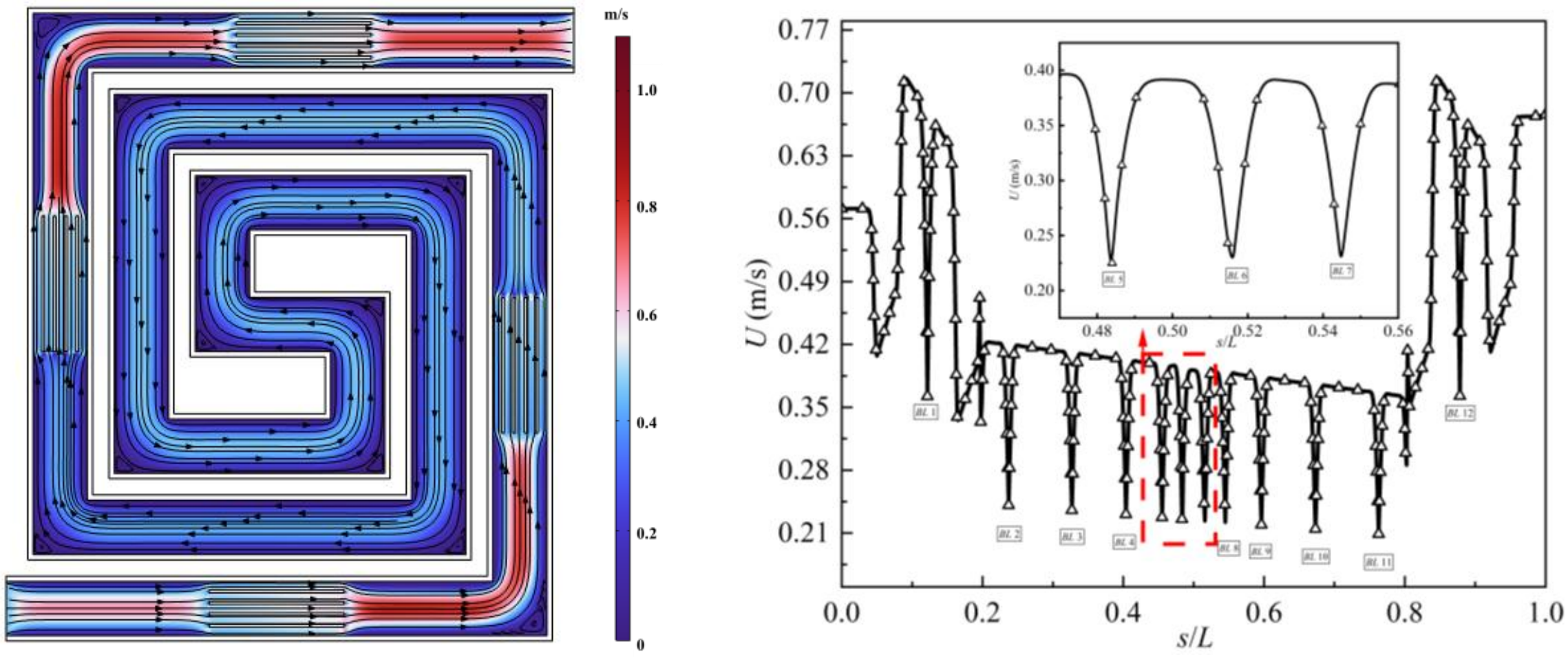

(a) Contour plot of velocity distribution. (b) Velocity distribution along the *L*-line.

Fig. 7. Velocity distribution characteristics under Case 1 condition.

To compare the velocity distribution within the microchannels at the same location and across different locations, Fig. 8 presents the velocity profile along the height ($h$) of the microchannel at Location 1 for Case 1, as well as the velocity comparison between different stages. At the same location, the velocity within the microchannels exhibits a similar symmetric distribution. For instance, in the symmetrically located *ML* 1-2 and *ML* 1-5, the velocity decreases from 0.51 m/s to approximately 0.41 m/s. In the symmetrically located *ML* 1-3 and *ML* 1-4, the velocity decreases from 0.52 m/s to approximately 0.44 m/s. On the other hand, when comparing the velocities between different locations, the velocity trends at Location 1 and Location 10 are similar. However, the velocities within the microchannels at Location 2 and Location 9 exhibit slight differences. The minimum velocity at *ML* 2-1 is 0.37 m/s, while the minimum velocity at *ML* 9-1 is 0.39 m/s. This difference can be attributed to the larger pressure differential at the microchannel ends at Location 2, which enhances the Poiseuille flow effect. This effect, in turn, inhibits the forward transport of gas.

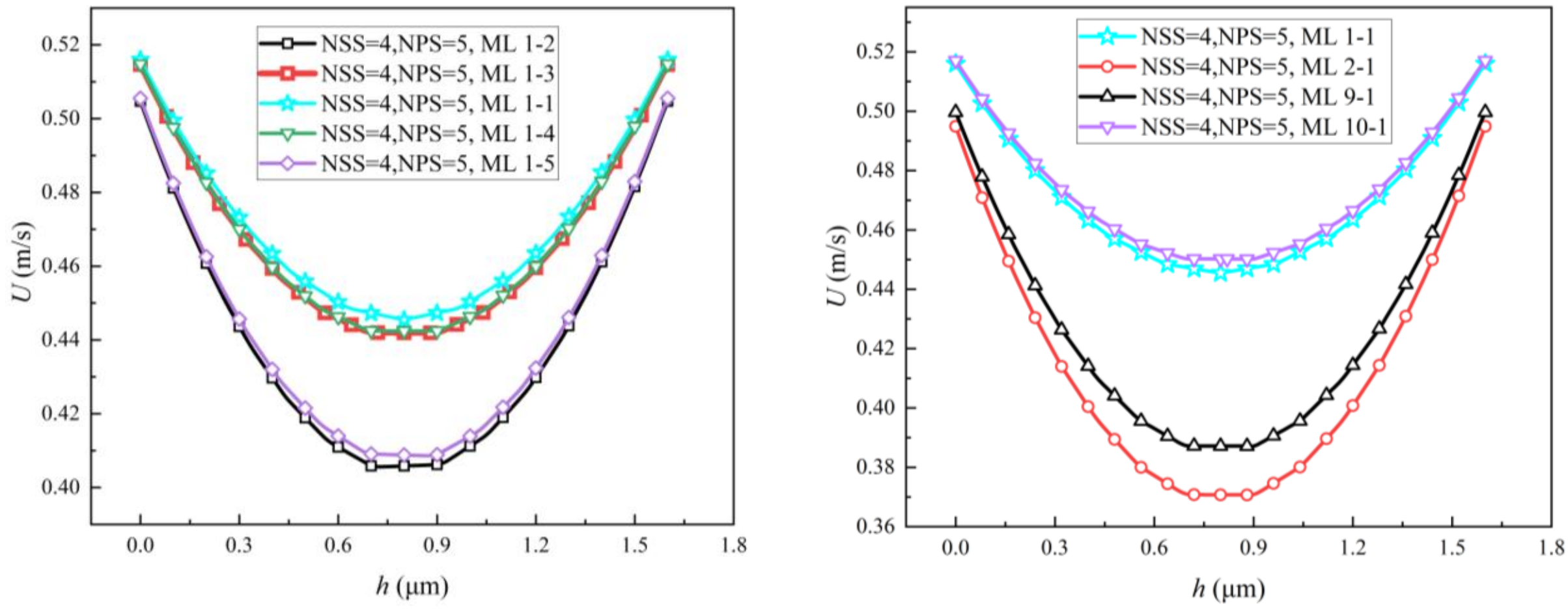

(a) Comparison of velocity along the channel height. (b) Velocity comparison between different stages.

Fig. 8. Velocity comparison within different microchannels at Location 1 and between different series stages under Case 1 condition.

### 3.2 Impact of the number of series microchannels on flow and transmission behaviors

To explore the impact of the number of series microchannels (NSS) on the flow behavior, Fig. 9 illustrates the velocity distribution within the Knudsen pump for different NSS values. It can be observed that as the NSS increases, the maximum velocity within the Knudsen pump gradually increases. For example, when NSS = 6, 8, and 10, the maximum velocities are 0.74 m/s, 0.93 m/s, and 0.97 m/s, respectively. This is mainly because, as the number of serial stages increases, flow separation begins to occur on the inner side of the main channel, leading to the formation of flow-separation vortices. After the gas from the branch microchannel enters the main channel, it immediately encounters a bend. At a finite Reynolds number, the main flow possesses sufficient inertia that it cannot instantaneously change direction and remain attached to the wall while undergoing a sharp 90° turn, causing the streamlines to separate at the inner convex corner. A low-velocity region consequently forms between the separated streamline and the downstream wall. Through viscous shear and entrainment, the separated shear layer drives reverse flow within this region, eventually forming a closed local recirculation zone, referred to as a flow-separation vortex [40]. This phenomenon reduces the effective flow cross-section, which ultimately causes an increase in the maximum flow velocity. This observation is fully consistent with the principle of mass conservation within the Knudsen pump. In contrast, in $L12$ and $L13$, the velocity exhibits little change for NSS values of 6, 8, and 10. This is because no additional microchannels are added in these main channels.

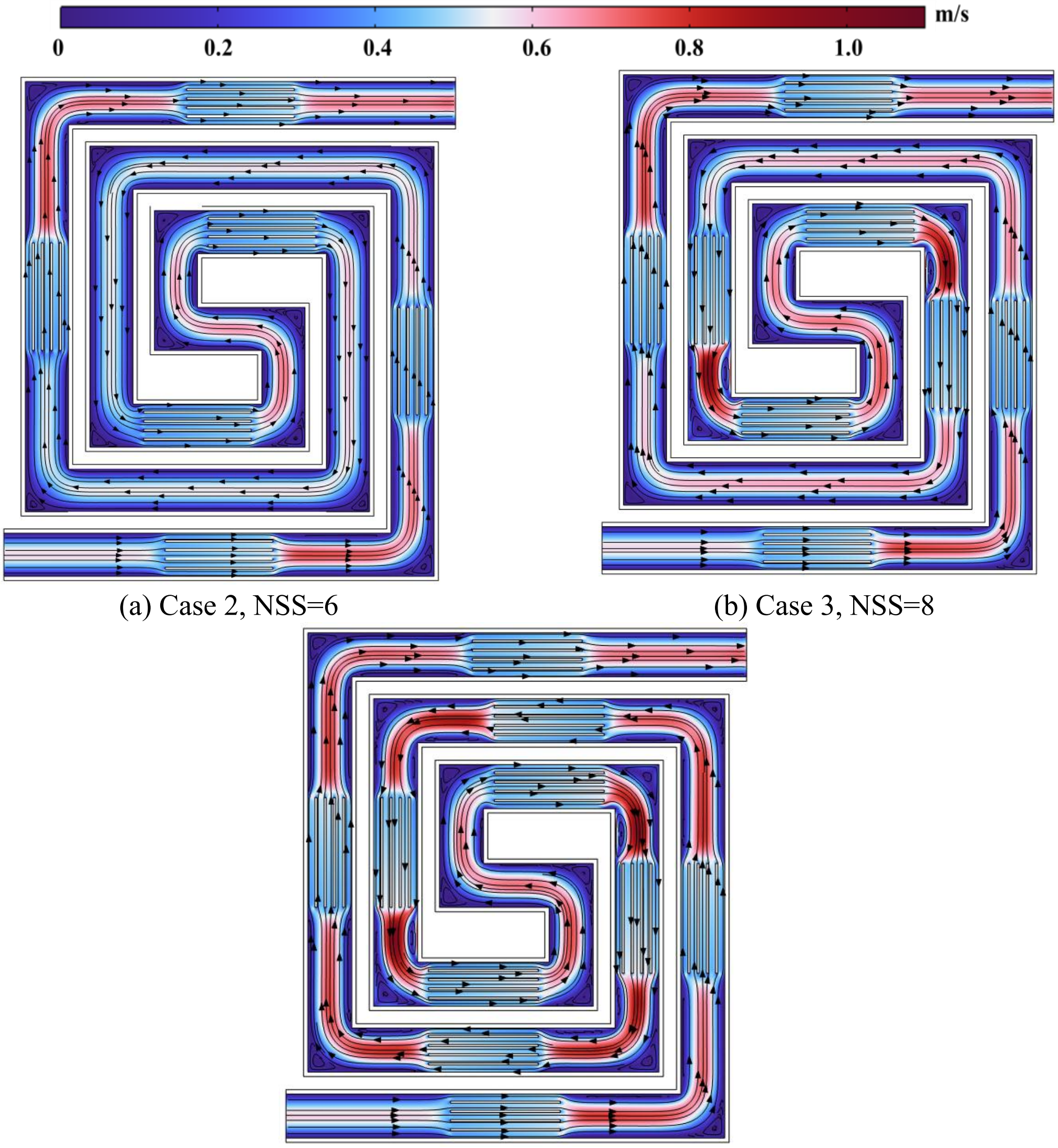

(a) Case 2, NSS=6 (b) Case 3, NSS=8

(c) Case 4, NSS=10

Fig. 9. Contour plot of velocity within the Knudsen pump for different NSS.

To investigate the impact of NSS on the transmission performance, Fig. 10 illustrates the changes in velocity distribution along the microchannel height at the $ML$ 1-1 position and the transmission performance of the Knudsen pump. For different numbers of series microchannels, the velocity distribution exhibits a parabolic profile. This arises from the superposition of the forward thermal transpiration flow and the backward Poiseuille flow driven by the pressure gradient. However, as the number of series microchannels increases, the minimum velocity gradually increases. For example, when the number of series stages is 6, 8, and 10, the minimum velocities are 0.458 m/s, 0.461 m/s, and 0.466 m/s, respectively, with average velocities of 0.519 m/s, 0.520 m/s, and 0.522 m/s, respectively, as shown in Fig. 10(a). This indicates that the increase in the number of series stages has a limited effect on the improvement of the flow behaviors within the microchannels. The transmission capacities for different series stages are $3.00 \times 10^{-7}$ kg·m$^{-1}$·s$^{-1}$ (NSS=4), $3.06 \times 10^{-7}$ kg·m$^{-1}$·s$^{-1}$ (NSS=6), $3.08 \times 10^{-7}$ kg·m$^{-1}$·s$^{-1}$ (NSS=8), and $3.10 \times 10^{-7}$ kg·m$^{-1}$·s$^{-1}$ (NSS=10). From the increase from 4 to 10 serial stages, the transmission performance only increases by about 3.35%. After conversion to the experimental volumetric flow rate ($Q_{sccm}$), the volumetric flow rates per unit depth are 25.04 sccm·mm$^{-2}$ (NSS = 4), 25.55 sccm·mm$^{-2}$ (NSS = 6), 25.70 sccm·mm$^{-2}$ (NSS = 8), and 25.85 sccm·mm$^{-2}$ (NSS = 10), respectively. The primary reason is that, as the number of serial stages increases, local pressure loads develop at the bends of the main channel, inducing disturbances in the flow field and consequently affecting the gas transmission performance. Therefore, the variation in pressure distribution at the bends constitutes a key factor governing the performance of multi-stage Knudsen pumps.

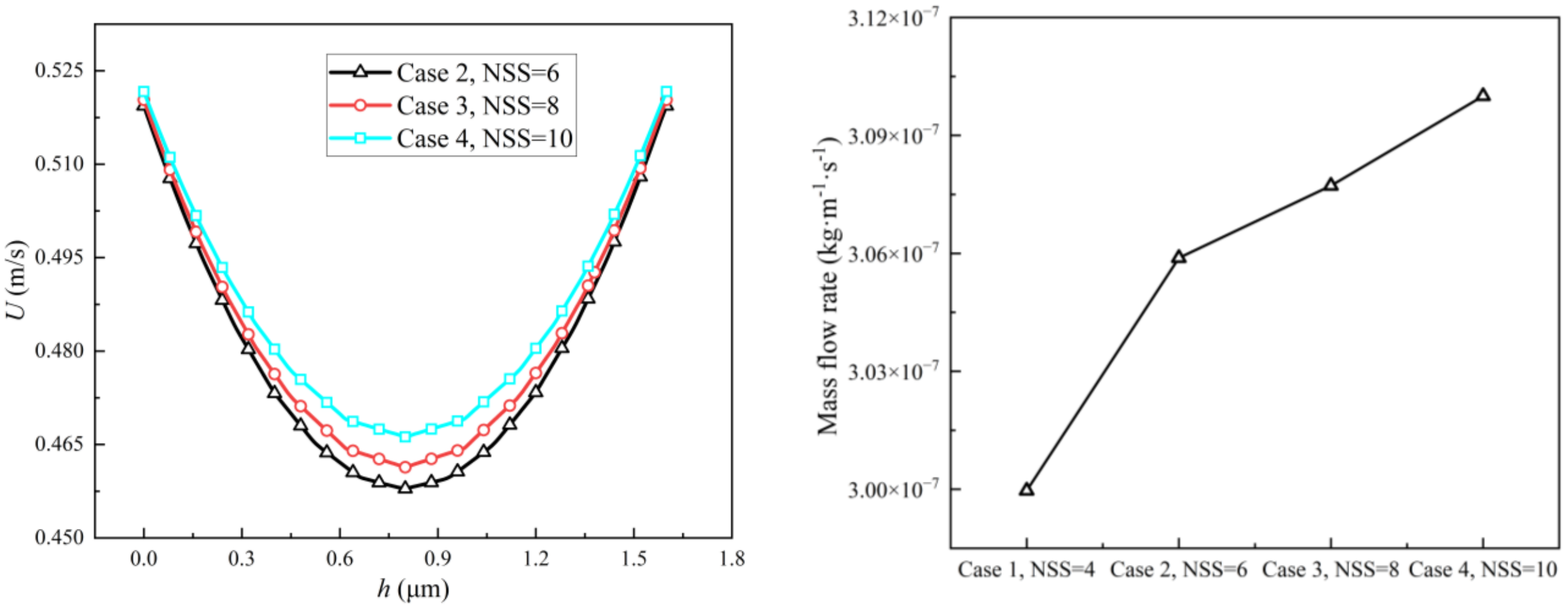

(a) Velocity profile of ML 1-1.

(b) Transmission performance of the Knudsen pump under different NSS.

Fig. 10. Velocity profile of ML 1-1 and mass flow rate for different NSS.

To further analyze the factors influencing the flow behaviors and transmission capacity, Fig. 11 presents the temperature distribution within the Knudsen pump for different numbers of series stages. The increase in the number of serial stages has a minimal impact on the overall temperature distribution, primarily manifested by the formation of new temperature gradients at the locations of the added microchannels. For example, compared to the *NSS* = 6 condition, in the *NSS* = 8 case, two sets of microchannels are added at Location 4 and Location 7. This results in changes in the temperature distribution at these two locations, where the temperature rises from 273 K at the inlet of the microchannels to 323 K at the outlet. However, the temperature changes at other locations are minimal. Therefore, the differences in temperature distribution are not the cause of the variations in the velocity distribution.

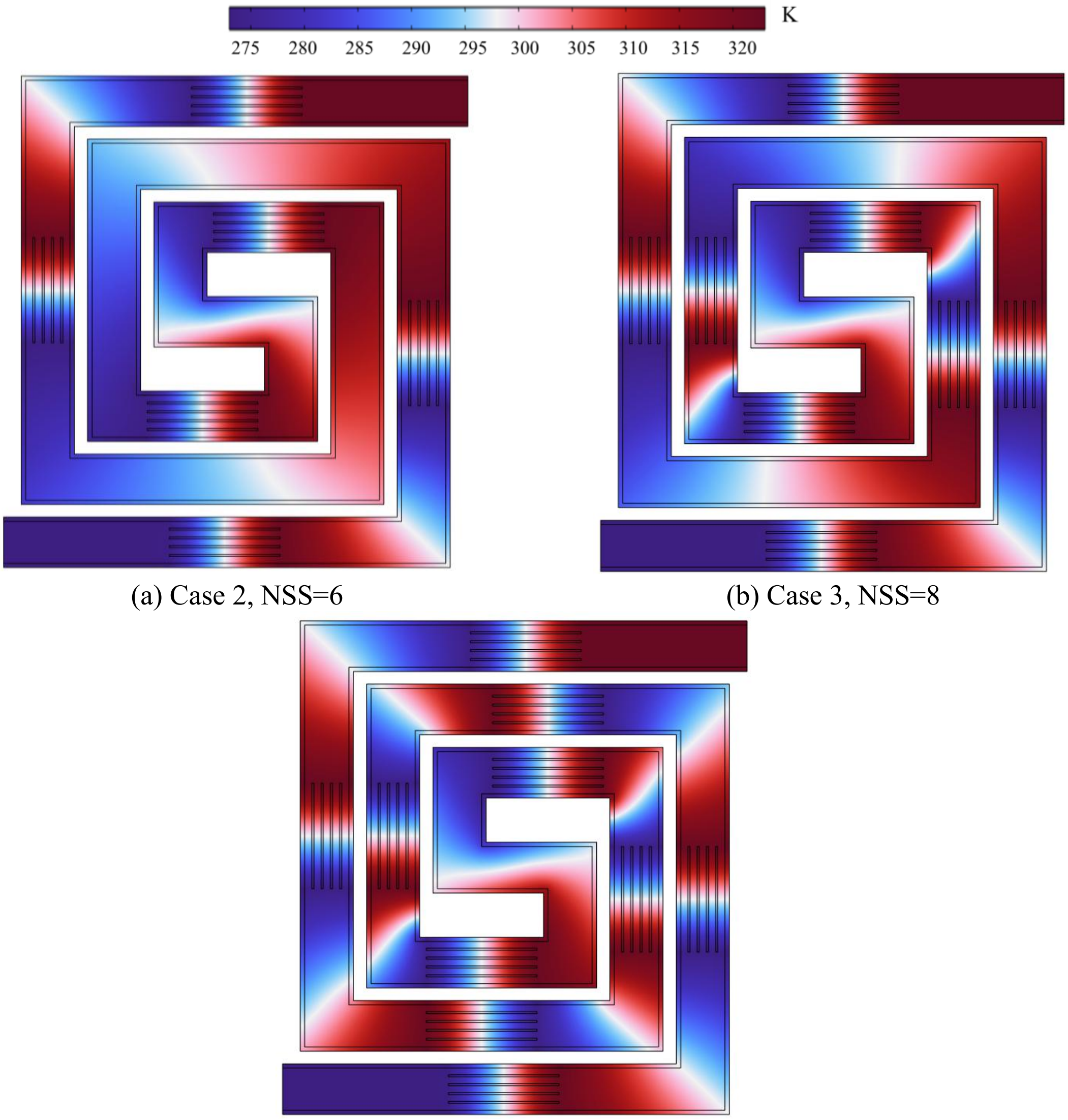

(a) Case 2, NSS=6

(b) Case 3, NSS=8

(c) Case 4, NSS=10

Fig. 11. Contour plot of temperature within the Knudsen pump for different NSS.

To further analyze the factors influencing the flow behaviors and transmission performance, Fig. 12 shows the Knudsen number and pressure distribution along $L1$ for different numbers of series microchannels. As the number of serial stages increases, the variation trend of the Knudsen number along the microchannel at L1 remains similar. Within the main channel, the Knudsen number is approximately 0.010, whereas in the microchannels it increases from approximately 0.064 at the inlet to approximately 0.078 at the outlet. The variation of the Knudsen number within the microchannels is relatively small. This is primarily because the mean free path in Eq. (11) is calculated based on the absolute pressure, and the pressure change induced by the increase in the number of microchannel stages is negligible compared to the absolute pressure itself. Consequently, its effect on the Knudsen number is limited. This indicates that the degree of rarefaction is not the cause of the differences in flow behaviors. On the other hand, as the number of series stages increases, the pressure difference across the microchannels gradually decreases. For example, when $NSS$ =

6, the pressure difference across the microchannels is 41 Pa. When *NSS* = 10, the pressure difference drops to 37 Pa. Therefore, the variation in pressure difference influences the Poiseuille effect, which in turn alters the flow resistance. When the pressure difference is large, the Poiseuille effect causes a more significant reduction in velocity within the microchannel, leading to a decrease in average velocity and resulting in differences in flow and transmission behaviors. This further indicates that increasing the number of serial stages can reduce the pressure load at both ends of the microchannel, thereby decreasing flow resistance. This, in turn, slightly enhances the transmission performance.

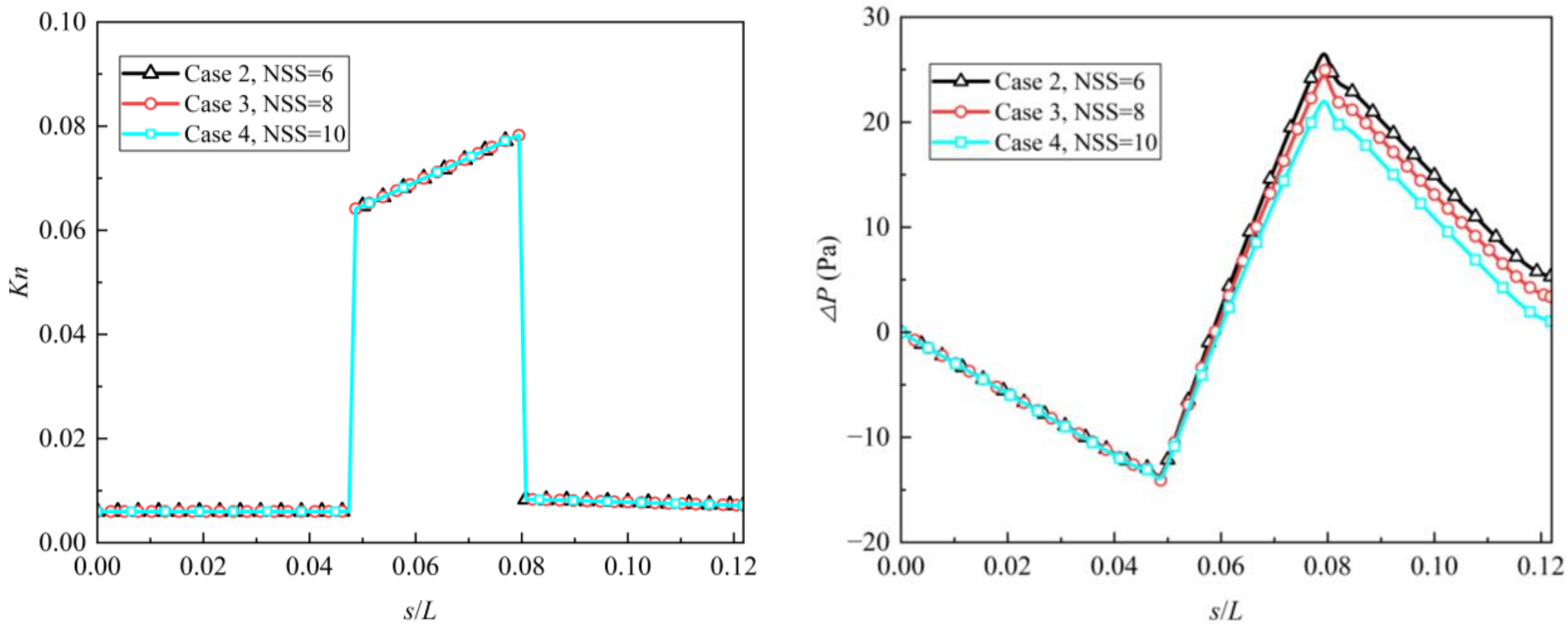

(a) Distribution of Knudsen number along $L1$. (b) Distribution of pressure along $L1$.

Fig. 12. Distribution of Knudsen number and pressure along $L1$ for different NSS.

### 3.3 Impact of the number of parallel microchannels on flow and transmission behaviors

To investigate the effect of NPS on the flow behavior of the Knudsen pump under a constant main channel size, Fig. 13 presents the velocity distributions within the Knudsen pump for different NPS configurations. As shown in Fig. 13, the maximum velocity within the Knudsen pump gradually increases with increasing numbers of parallel stages. For example, when NPS = 2, 3, 4, and 5, the maximum velocities are 0.40 m/s, 0.62 m/s, 0.71 m/s, and 0.73 m/s, respectively. This indicates that increasing the number of parallel stages can effectively enhance the flow velocity. However, unlike the increase in the number of serial stages, which induces local recirculation and the formation of flow-separation vortices on the inner side of the main channel, increasing the number of parallel microchannels does not induce additional flow-separation vortices on the inner side of the main channel along the flow direction, nor does it reduce the effective flow cross-sectional area. On the other hand, similar to the case of increasing the number of serial stages, Moffatt vortices still form at the outer concave corners of each bend in the main channel as the number of parallel microchannels increases. These vortices can negatively affect the flow continuity and stability.

To address this issue, further investigation into the impact of the bent channel structure on flow behaviors is necessary. For example, the flow can be improved by optimizing the structure at the bend locations, such as using rounded corners on both the inner and outer sides of the bend, which would help reduce the negative impact of flow vortices on flow stability.

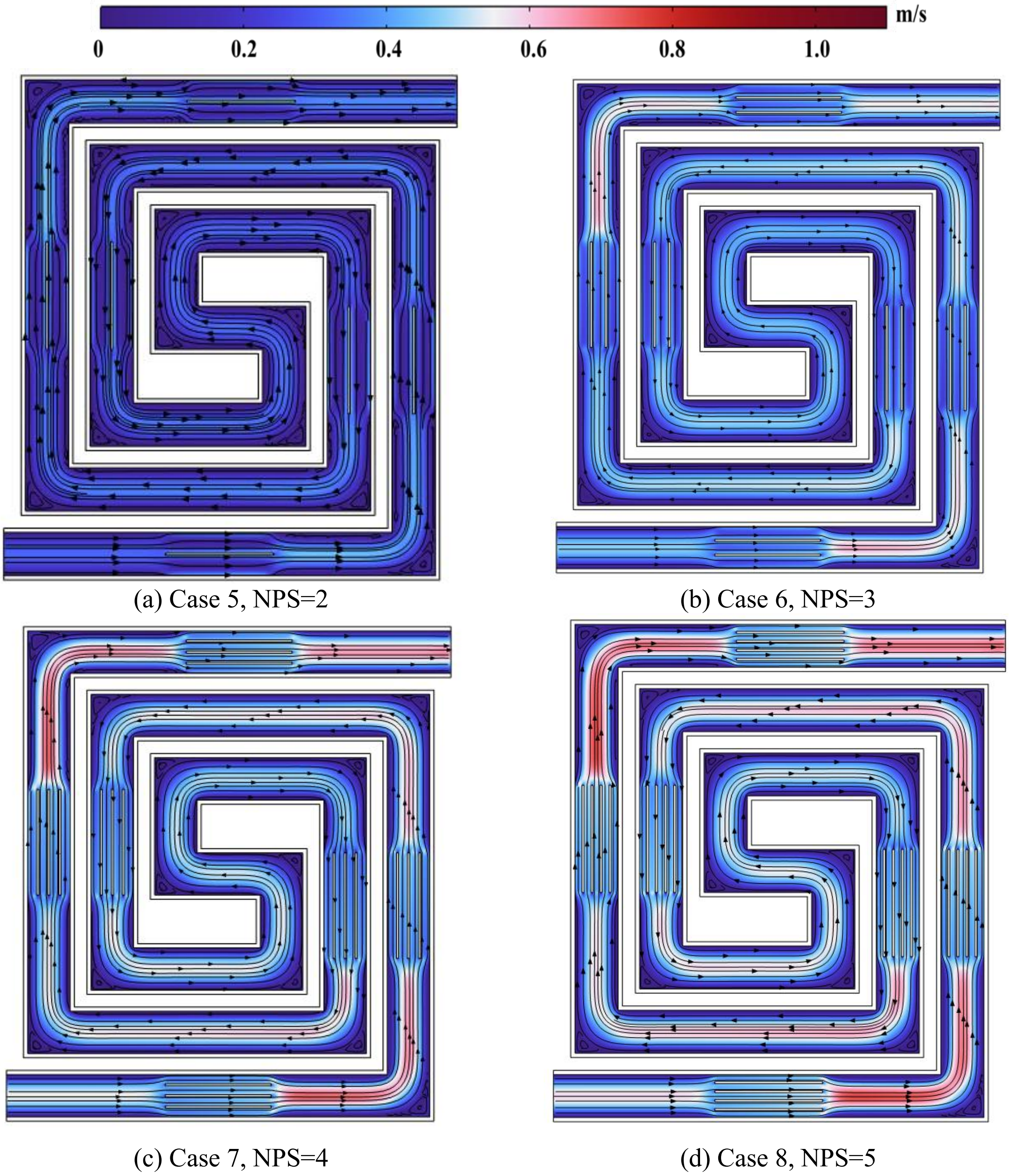

Fig. 13. Contour plot of velocity within the Knudsen pump for different NPS.

To investigate the impact of parallel microchannels on the transmission capability, Fig. 14 shows the velocity profile along ML 1-1 and the corresponding mass flow rate. As the number of parallel stages increases, the slip velocity at the microchannel boundaries gradually increases, and the velocity distribution for different parallel stages follows a parabolic pattern. For example, when NPS = 2, 3, 4, and 5, the slip velocities at the microchannel boundaries are 0.49 m/s, 0.50 m/s, 0.51 m/s, and 0.52 m/s, respectively, while the average velocities are 0.23 m/s, 0.39 m/s, 0.45 m/s, and 0.48 m/s. This indicates that the increase in the number of parallel stages enhances both the slip velocity at the boundary and the average velocity within the microchannels, as shown in Fig. 14(a). Correspondingly, the mass flow rates for different NPS are 1.80 × $10^{-7}$ kg·m$^{-1}$·s$^{-1}$ (NPS=2), 2.67 × $10^{-7}$ kg·m$^{-1}$·s$^{-1}$ (NPS=3), 3.00 × $10^{-7}$ kg·m$^{-1}$·s$^{-1}$ (NPS=4), and 3.07 × $10^{-7}$ kg·m$^{-1}$·s$^{-1}$ (NPS=5), as shown in Fig. 14(b). Furthermore, we observe that as the number of parallel microchannels increases, the rate of increase in transmission performance gradually decreases. For instance, when the number of parallel stages increases from 2 to 3, the transmission performance increases by approximately 48.9%. However, when the number of parallel stages increases from 4 to 5, the transmission performance only increases by about 2.3%. Overall, the transmission performance increases by 70.80% when the number of parallel stages increases from 2 to 5. After conversion to the experimental volumetric flow rate ($Q_{sccm}$), the volumetric flow rates per unit depth are 12.64 sccm·mm$^{-2}$ (NPS=2), 19.78 sccm·mm$^{-2}$ (NPS=3), 23.60 sccm·mm$^{-2}$ (NPS=4), and 25.63 sccm·mm$^{-2}$ (NPS=5), respectively.

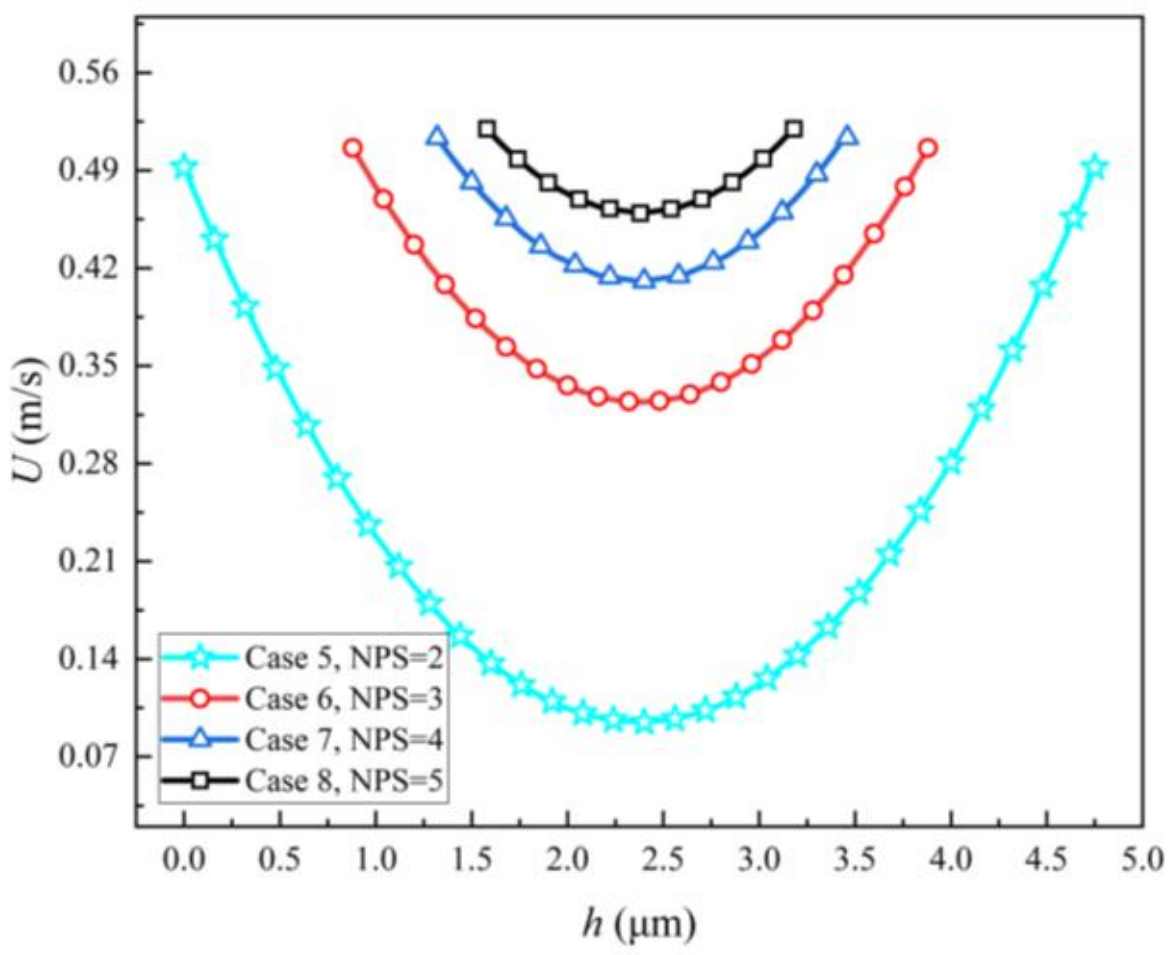

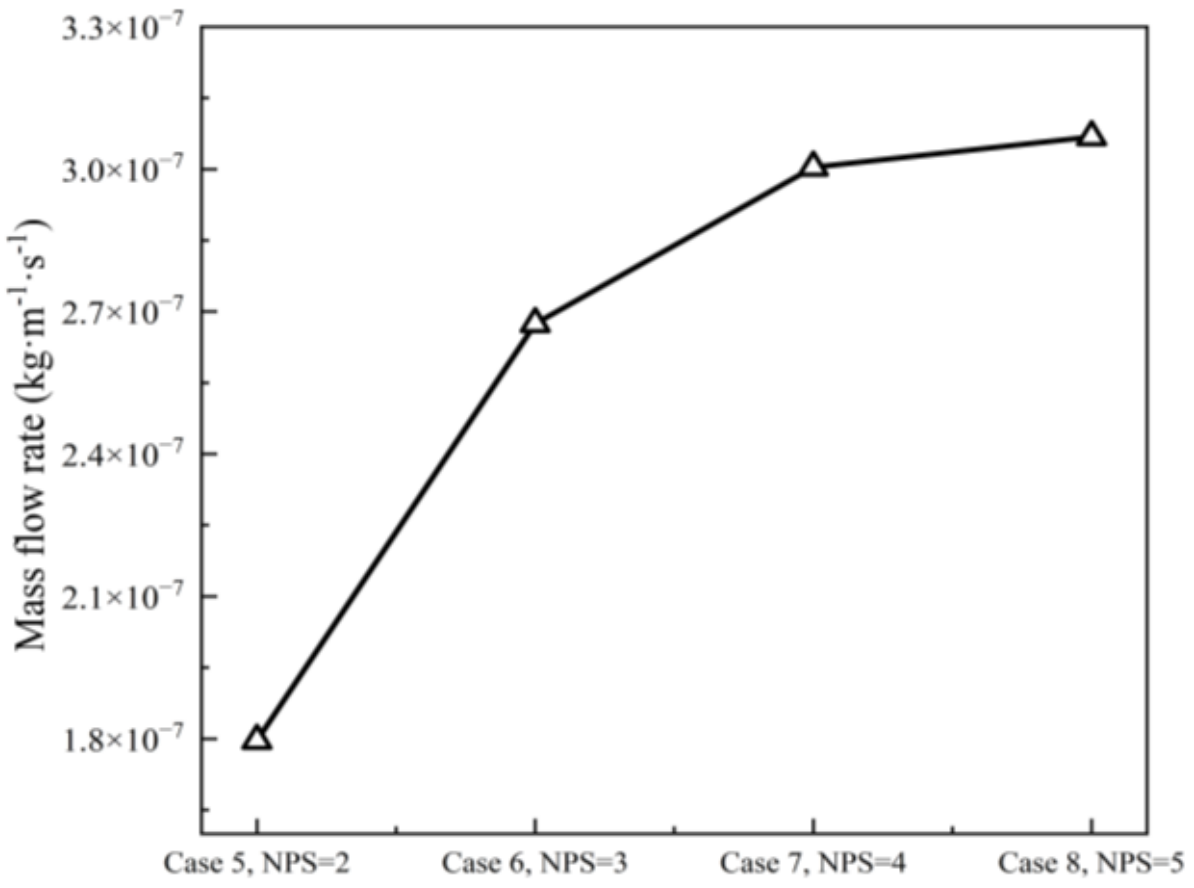

(a) Velocity profile along the height at Location 1. (b) Transmission performance for different NPS.

Fig. 14. Velocity profile of the central microchannel at Location 1 and mass flow rate for different NPS.

To analyze the factors influencing the flow and behaviors, Fig. 15 shows the temperature distribution within the Knudsen pump for different numbers of parallel stages. Similar to the effect of increasing serial stages on the temperature distribution, the increase in the number of parallel stages has a minimal impact on the overall temperature distribution. The temperature within the microchannels increases from 273 K at the inlet to 323 K at the outlet. However, within the main channel, the temperature decreases from 323 K at the microchannel outlet to 273 K at the inlet of the next microchannel. Therefore, the differences in temperature distribution are not the cause of the differences in velocity distribution.

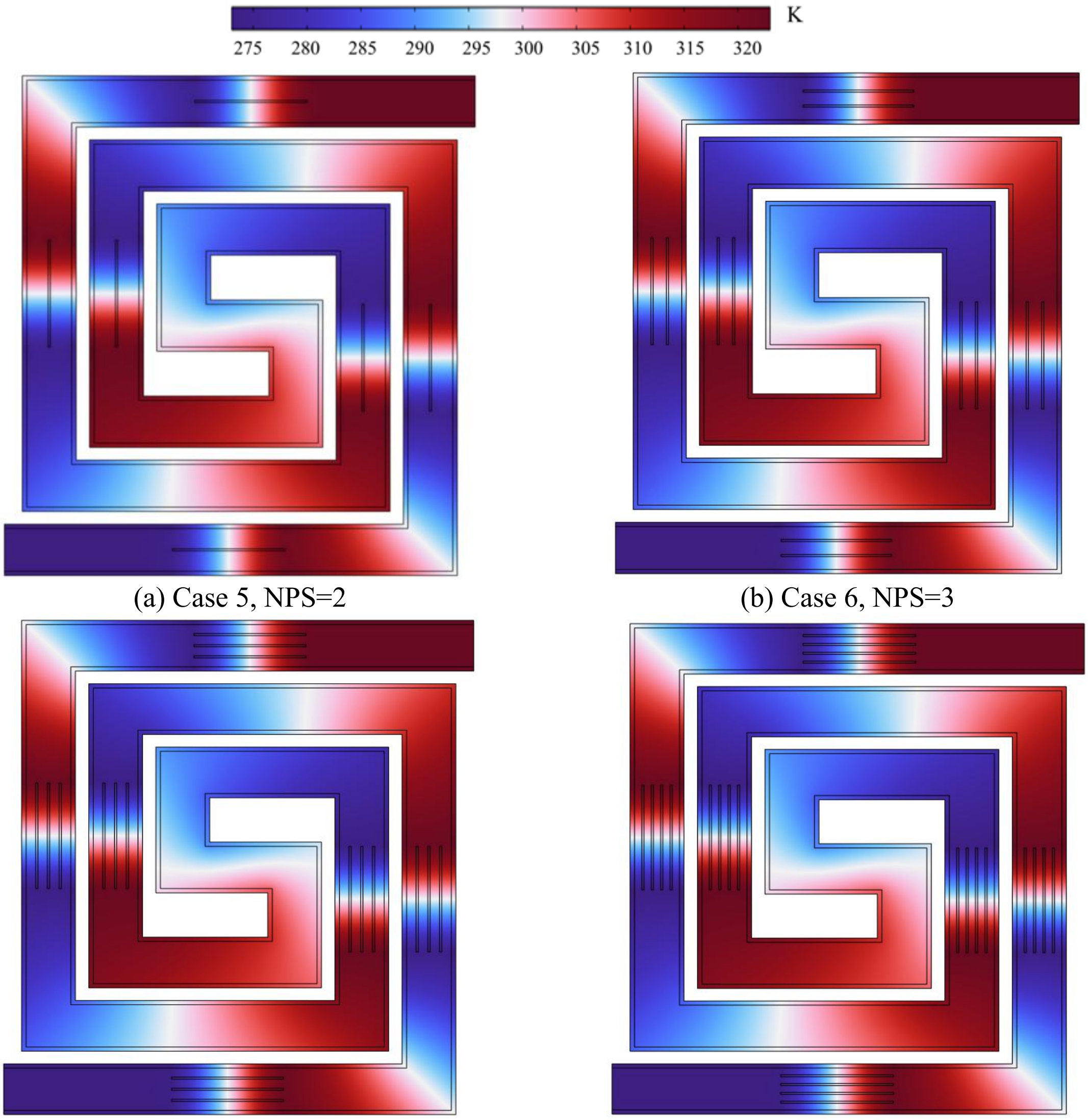

(a) Case 5, NPS=2 (b) Case 6, NPS=3

(c) Case 7, NPS=4 (d) Case 8, NPS=5

Fig. 15. Contour plot of temperature within the Knudsen pump for different NPS.

To further explore the reasons behind the velocity differences within the microchannels, Fig. 16 shows the pressure and Knudsen number distributions for different NPS. From the Knudsen number distribution, shown in Fig. 16(a), it can be observed that as the number of parallel microchannels increases, the Knudsen number within the microchannels gradually increases. For example, when NPS = 2, the Knudsen number ranges from 0.018 to 0.023. When NPS = 5, the range is from 0.064 to 0.078. This suggests that as the number of parallel microchannels increases, the rarefaction effect within the microchannels becomes more pronounced. On the other hand, as the number of parallel microchannels increases, the pressure difference across the microchannels gradually increases. When NPS = 2, 3, 4, and 5, the pressure differences are 30 Pa, 35 Pa, 39 Pa, and 40 Pa, respectively, as shown in Fig. 16(b). This indicates that as the number of parallel microchannels increases, the pressure difference between the inlet and outlet of the microchannels gradually rises, imposing a greater pressure load. The amplified pressure difference strengthens the backward Poiseuille flow. As a result, the rate of increase in transmission performance gradually diminishes.

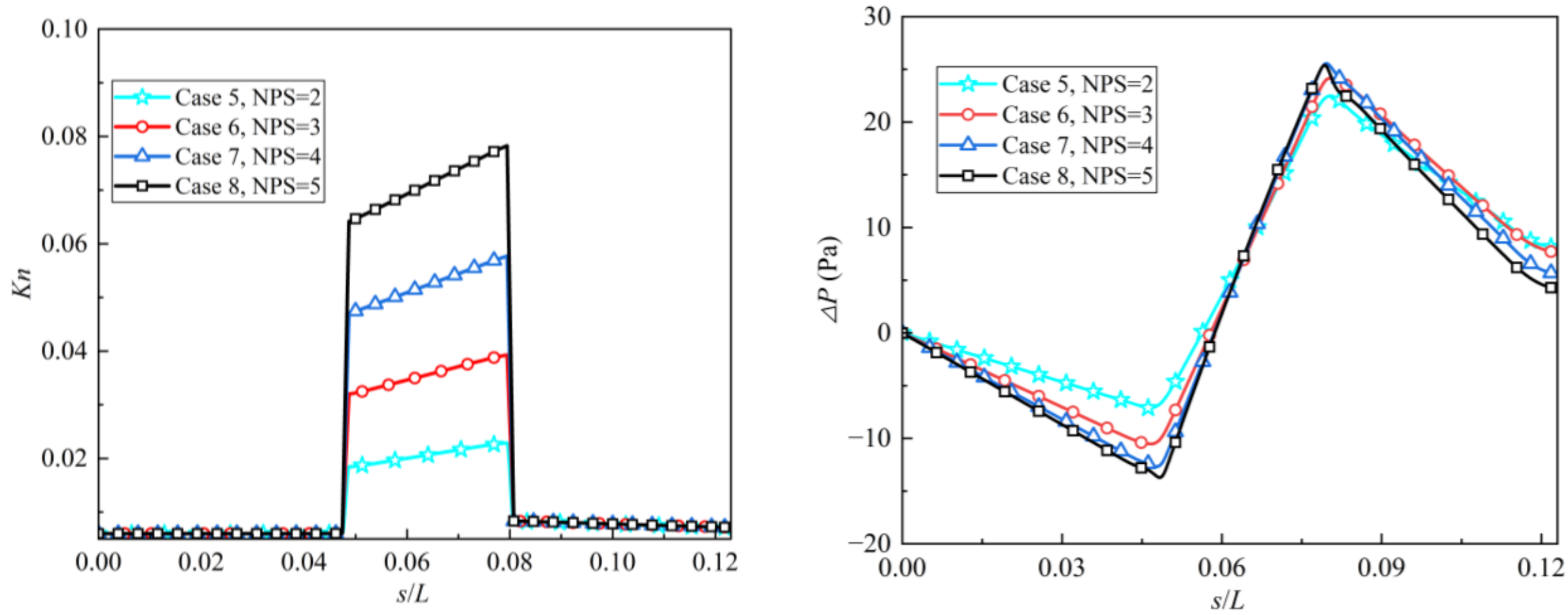

(a) Knudsen number distribution. (b) Pressure distribution.

Fig. 16. Pressure and Knudsen number distributions of the central microchannel at Location 1 for different NPS.

On the other hand, the effects of the number of parallel microchannels on the flow and transmission behaviors under a constant microchannel characteristic dimension are further analyzed, as shown in Figs. S2–S5 of the Supporting Information. As NPS increases, the suppressive effect of the main channel on the flow through the microchannels gradually becomes stronger, resulting in a progressively reduced rate of improvement in the overall transmission performance and causing it to deviate from a linear increasing trend. Specifically, at NPS values of 2, 3, 4, and 5, the transmission performance of a single microchannel is $6.66 \times 10^{-8}$ $kg \cdot m^{-1} \cdot s^{-1}$, $6.55 \times 10^{-8}$ $kg \cdot m^{-1} \cdot s^{-1}$, $6.36 \times 10^{-8}$ $kg \cdot m^{-1} \cdot s^{-1}$, and $6.14 \times 10^{-8}$ $kg \cdot m^{-1} \cdot s^{-1}$, respectively. Further analysis of the pressure distributions shows that, at NPS = 2 and 3, the pressure difference across the microchannel remains nearly unchanged, with a variation of less than 1 Pa. However, when NPS increases to 4 and 5, the pressure difference increases to 29 Pa and 40 Pa, respectively. These results indicate that, when the container-to-microchannel flow cross-sectional area ratio falls below 10:6.4, the pressure loading induced by the container increases significantly. The increased pressure loading strengthens the reverse Poiseuille flow and consequently reduces the transmission performance of individual microchannels. This result is consistent with Li et al. [41], who reported that end containers induce an additional pressure difference across the microchannels, thereby generating reverse flow resistance and suppressing gas transport.

# 4. Discussion

## 4.1 Comparison of transmission performance of Knudsen Pumps

In terms of transmission performance, the hydrogen Knudsen pump proposed in the present study (Case 4) achieves a maximum standard volumetric flow rate per unit effective opening area of 25.85 sccm·mm$^{-2}$ ($3.10 \times 10^{-7}$ kg·m$^{-1}$·s$^{-1}$). In comparison, the serpentine multi-stage hydrogen Knudsen pump with alternating curved and straight channels proposed by Lan et al., which employs seven parallel microchannels with a total effective outlet opening width of 14 μm, achieves a maximum mass flow rate per unit depth of $1.12 \times 10^{-7}$ kg·m$^{-1}$·s$^{-1}$ [42]. After normalization by the outlet height, its maximum standard volumetric flow rate per unit effective opening area is approximately 5.34 sccm·mm$^{-2}$. Therefore, the standard volumetric flow rate per unit effective opening area of the Knudsen pump proposed in the present study is approximately 4.84 times that of the serpentine multi-stage hydrogen Knudsen pump with alternating curved and straight channels. In addition, Bond et al. numerically investigated eight-stage Knudsen pumps with alternating curved and straight channels, double-curved channels, and sinusoidal channels under a temperature difference of 600 K using argon as the working gas [20,21]. The results showed that the maximum mass flow rates per unit depth of the alternating curved–straight and double-curved configurations were $2.4 \times 10^{-6}$ kg·m$^{-1}$·s$^{-1}$ ($Kn$ = 0.38) and $6.4 \times 10^{-6}$ kg·m$^{-1}$·s$^{-1}$ ($Kn$ = 0.24), respectively. For the sinusoidal configuration, the maximum dimensionless mass flow rate was 0.023 at $Kn$ = 0.39, while the corresponding dimensional mass flow rate was not reported. Based on the reported transmission capacity of the sinusoidal configuration, which is 240% of that of the double curved configuration, its maximum mass flow rate per unit depth can be estimated as approximately$1.536 \times 10^{-5}$ kg·m$^{-1}$·s$^{-1}$ [20]. After normalization by the temperature difference, the transmission capacities of the three configurations are $4.0 \times 10^{-9}$ kg·m$^{-1}$·s$^{-1}$·K$^{-1}$, $1.07 \times 10^{-8}$ kg·m$^{-1}$·s$^{-1}$·K$^{-1}$, and $2.56 \times 10^{-8}$ kg·m$^{-1}$·s$^{-1}$·K$^{-1}$, respectively. In comparison, the novel Knudsen pump proposed in the present study (Case 4), operating with argon under a temperature difference of 50 K, achieves a maximum mass flow rate per unit depth of $1.4 \times 10^{-6}$ kg·m$^{-1}$·s$^{-1}$. Its transmission capacity per unit temperature difference is $2.8 \times 10^{-8}$ kg·m$^{-1}$·s$^{-1}$·K$^{-1}$. Compared with the three configurations investigated by Bond et al., the normalized transmission capacity of the configuration proposed in the present study is 7.00, 2.62, and 1.09 times the corresponding values for the alternating curved–straight, double curved, and sinusoidal configurations, respectively.

On the other hand, experimental studies on hydrogen Knudsen pumps remain relatively limited, and existing experiments have mainly employed air as the working gas. For example, in 2021 [8], 2022 [7], and 2024 [9], Byambadorj et al. fabricated various silicon-based Knudsen pumps using a six-mask micromachining process. The operating temperature differences considered in their numerical simulations were 84 K, 36 K, and 102 K, respectively. The corresponding maximum standard volumetric flow rates per unit effective opening area were 16.5 sccm·mm$^{-2}$, 8.2 sccm·mm$^{-2}$, and 42.7 sccm·mm$^{-2}$, respectively. The maximum transmission performances per unit temperature difference were 0.196 sccm·mm$^{-2}$·K$^{-1}$, 0.228 sccm·mm$^{-2}$·K$^{-1}$, and 0.419 sccm·mm$^{-2}$·K$^{-1}$, respectively. For comparison, when the Knudsen pump proposed in the present study operates with air under a temperature difference of 50 K, its maximum standard volumetric flow rate per unit effective opening area is 14.01 sccm·mm$^{-2}$. The corresponding maximum transmission performance and pressure-generation capability per unit temperature difference is 0.280 sccm·mm$^{-2}$·K$^{-1}$. It should be noted that the operating gases, microchannel lengths, temperature differences, numbers of stages, and degrees of rarefaction adopted in these studies are not identical. Therefore, the above comparisons are primarily intended to illustrate the performance ranges of different Knudsen pump configurations rather than to provide a rigorous comparison under identical operating conditions.

### 4.2 Analysis of the degree of integration

In terms of chip integration, the single-stage serpentine Knudsen pump proposed by Lan et al. occupies a footprint of approximately (40 + 30) μm × (15 + 30) μm = 3150 μm$^2$, corresponding to a total footprint of approximately 22050 μm$^2$ for the seven-stage series configuration [42]. In comparison, the ten-stage spiral Knudsen pump proposed in the present study occupies only 100 μm × 100 μm = 10000 μm$^2$, approximately 45% of the footprint of the former, indicating a higher degree of integration. The Knudsen pumps reported by Byambadorj et al. in 2021, 2022, and 2024 have footprints of 5 mm × 7.5 mm = 37.5 mm$^2$ [8], 5.4 mm × 7.8 mm = 42.12 mm$^2$ [7], and 4.5 mm × 3.75 mm = 16.88 mm$^2$ [9], respectively, all substantially larger than that of the configuration proposed in the present study. Further considering structural scaling, if the overall dimensions of the spiral Knudsen pump are increased to 4 mm × 4 mm and seven main channels are arranged in the vertical direction, each main channel can occupy a width of approximately 0.57 mm. Under the ideal condition of maintaining a microchannel pitch of 1.6 μm, approximately 350 microchannels can be arranged along each main channel, which is about 70 times the number of microchannels in the current 10 μm wide main channel. Therefore, under the ideal assumption that all other structural parameters, boundary conditions, and the transmission performance of each microchannel remain unchanged, the total flow rate of the pump could theoretically increase to approximately 70 times that of the current configuration.

## 5. Conclusions

This study proposes a novel spiral-structured multi-stage Knudsen pump based on temperature-driven thermal transpiration and pressure-driven Poiseuille flow. The hydrogen transport mechanism in the Knudsen pump is investigated. On this basis, the effects of the numbers of parallel and serial microchannels on flow behavior and transmission performance are further analyzed. The main conclusions are as follows:

In the spiral-shaped Knudsen pump, the frequent bends in the main channel significantly alter the flow path, inducing pronounced changes in the magnitude and direction of velocity, which increase the complexity of the local flow field. As a result, nonlinear temperature distributions develop near the regions with frequent bends, reflecting the intricate interplay between flow dynamics and thermal transport. At these bends, the abrupt change in flow direction causes significant gas deceleration, accompanied by a slight local increase in static pressure. This indicates that the bend geometry significantly affects the local velocity and pressure distributions. In addition, the flow near the outer concave corners is dominated by viscous effects, leading to the formation of Moffatt vortices and further increasing the complexity of the local flow structure.

As the number of serial microchannel stages increases, flow separation occurs at the inner convex corners of the main channel due to inertial effects, leading to the formation of flow-separation vortices. Moreover, with the increase in the number of serial stages, the flow separation phenomenon intensifies. The appearance of flow vortices reduces the flow cross-section, increasing the velocity within the main channel. On the other hand, increasing the number of serial microchannels with temperature gradients weakens the inlet-to-outlet pressure difference induced by the reservoir effect, thereby suppressing the pressure-driven Poiseuille flow and reducing flow resistance. This change slightly increases the average flow velocity in the microchannels, thereby improving the transmission performance. When the number of serial microchannels reaches 10, the transmission performance improves by approximately 3.35%.

In contrast to the behavior observed with increasing serial microchannels, no flow separation occurs along the inner side of the flow direction within the main channel as the number of parallel microchannels increases. Vortices are formed only on the outer side of the flow direction. Furthermore, the increase in the number of parallel microchannels enhances the rarefaction effect within the microchannels, reducing molecular collision frequency and increasing the mean free path. This, in turn, amplifies the intensity of thermal transpiration and increases the pressure difference across the microchannel. As a result, transmission performance is gradually improved with the increase in the number of parallel stages. However, this improvement gradually slows down. When the number of parallel microchannels reaches 5, transmission performance increases by 70.80%.

## Supplementary materials

## S1. Effect of parallel microchannel number on flow and transmission behaviors at a constant microchannel characteristic dimension

### S1.1 Geometric models with a constant microchannel characteristic dimension

To investigate the effect of the number of parallel microchannels on transmission performance while maintaining a constant microchannel characteristic dimension, the 1.6 μm characteristic dimension of Case 8 with *NSS* = 6 was adopted as the basis for constructing the comparison cases, as shown in Fig. S1. The comparison models include cases with *NPS* = 2, 3, and 4. In all cases, the height of each microchannel was fixed at 1.6 μm to ensure the same degree of gas rarefaction. The microchannels were positioned at Locations 1, 2, 4, 7, 9, and 10, and the detailed configurations of the comparison cases are summarized in Table S1. For NPS = 2, 3, and 4, the container-to-microchannel flow cross-sectional area ratios are 10:3.2, 10:4.8, and 10:6.4, respectively, compared with 10:8 for Case 8.

Table S1 Comparison cases with different numbers of parallel microchannels at a constant microchannel characteristic dimension

| Case | Research issue | Number of Parallel Stages (NPS) | Microchannel Characteristic Dimension *h* (μm) | Number of Serial Stages (NSS) | Locations |
|---|---|---|---|---|---|
| Comparison Case 1 | Effects of parallel stages on flow and transmission behaviors | 2 | 1.6 | 6 | Locations 1, 2, 4, 7, 9, and 10 |
| Comparison Case 2 | | 3 | | | Locations 1, 2, 4, 7, 9, and 10 |
| Comparison Case 3 | | 4 | | | Locations 1, 2, 4, 7, 9, and 10 |

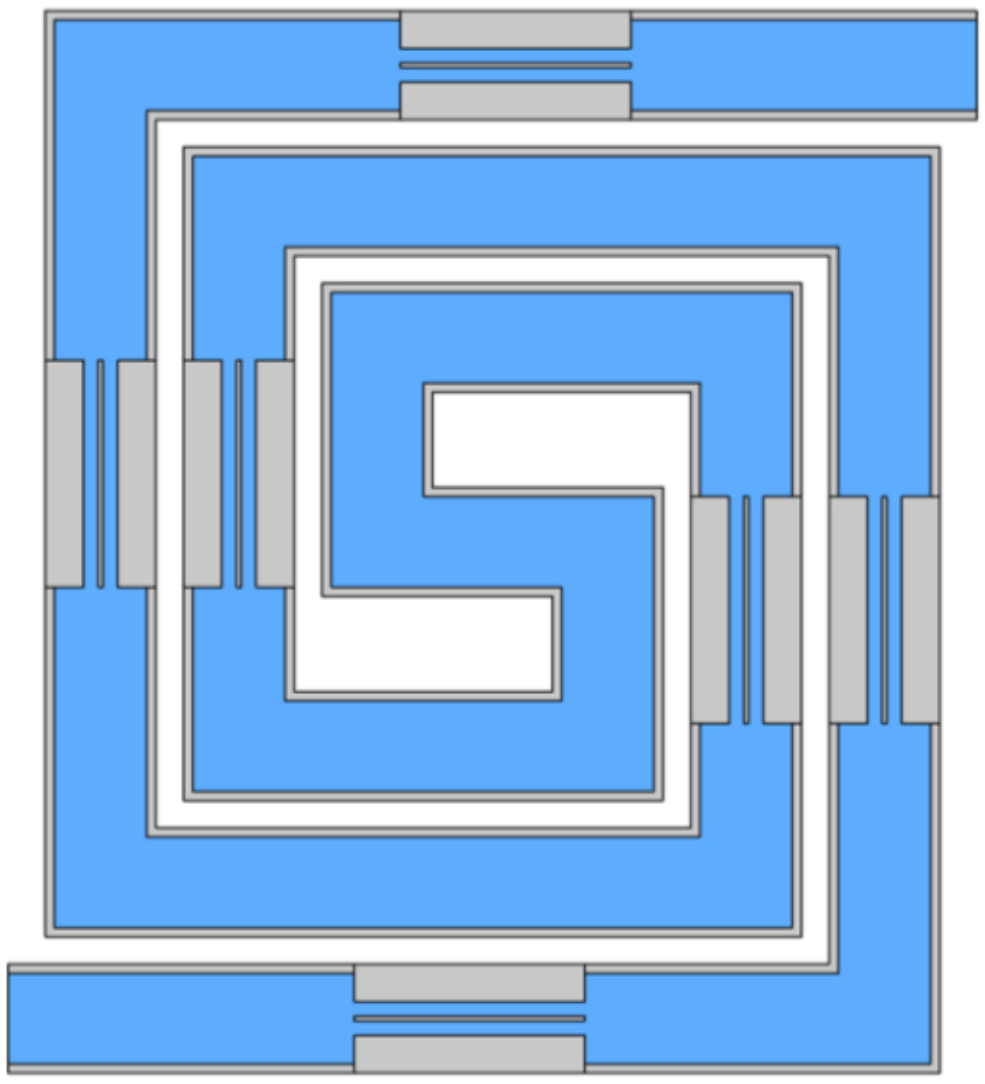

(a) NSS = 6, NPS = 2.

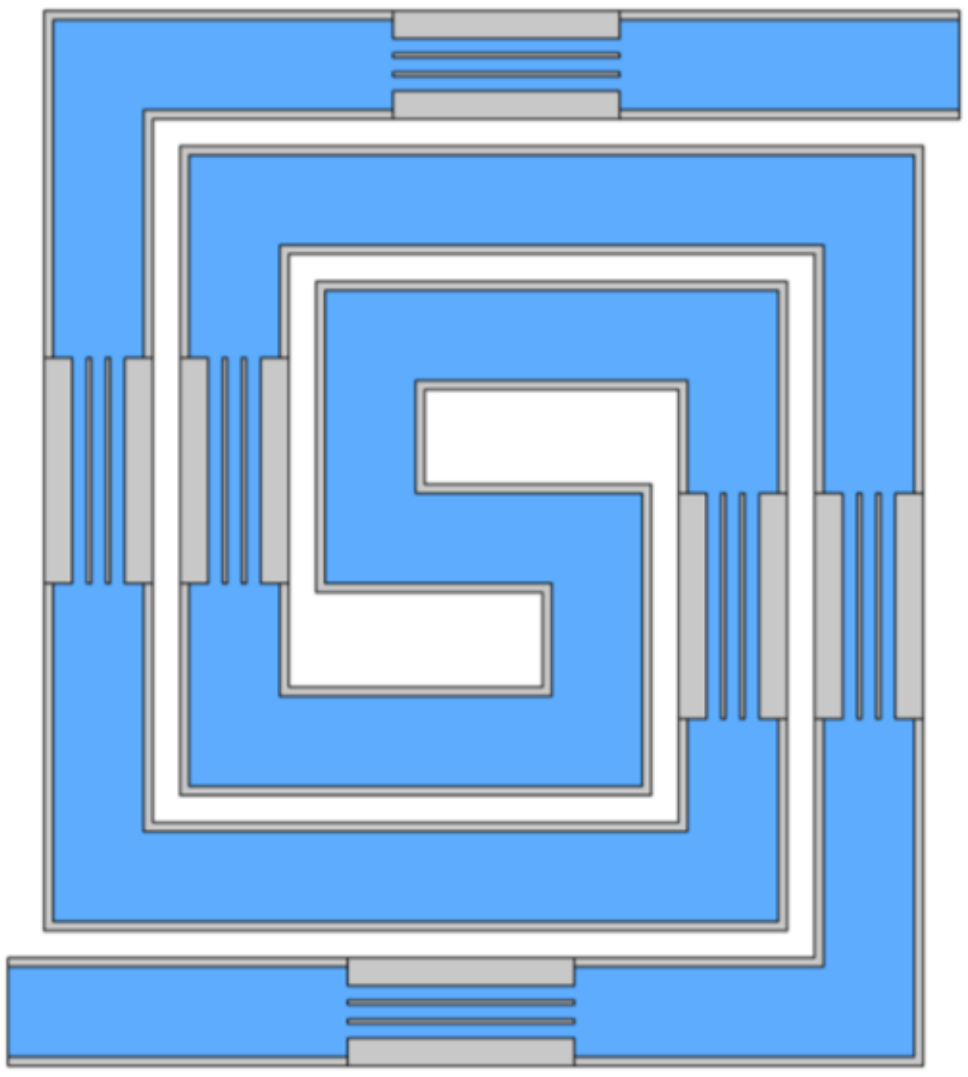

(b) NSS = 6, NPS = 3.

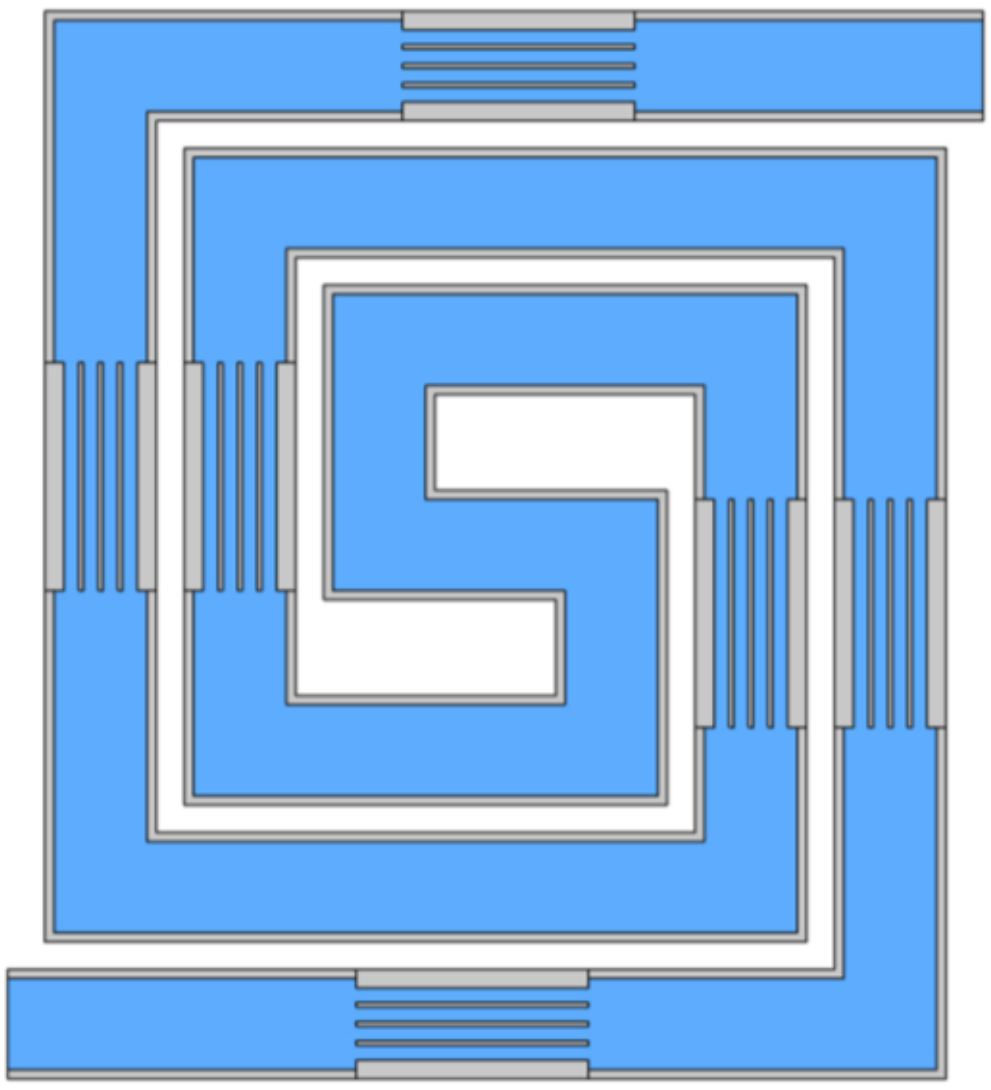

(c) NSS = 6, NPS = 4.

Fig. S1. Knudsen pump models with different numbers of parallel microchannels at a constant microchannel characteristic dimension.

**S1.2 Impact of the number of parallel microchannels on flow and transmission behaviors at a constant microchannel characteristic dimension**

To investigate the effect of the number of parallel stages (NPS) on the flow behavior of the Knudsen pump while maintaining a constant microchannel characteristic dimension, Fig. S2 presents the velocity contours within the Knudsen pump for different NPS. As NPS increases, the maximum velocity in the main channel gradually increases. For example, at $NPS$ = 4, the maximum velocity in the main channel reaches 0.55 m/s, indicating that increasing the number of parallel stages effectively enhances the gas flow velocity in the main channel. In addition, similar to the flow behaviors observed when the characteristic dimension of the main channel is kept constant, Moffatt vortices are still formed at the outer concave corners of the bends in the main channel as the number of parallel microchannels increases. However, the local flow behavior within the microchannels exhibits a different trend under the two conditions. As shown in Fig. S3(a), the velocity distributions within the microchannels are nearly identical for NPS = 2 and 3, with wall slip and centerline velocities of approximately 0.528 m/s and 0.493 m/s, respectively. When NPS increases to 4, the wall slip and centerline velocities decrease to 0.523 m/s and 0.459 m/s, respectively, with the centerline velocity decreasing by approximately 12.2%. The reduction in flow velocity within an individual microchannel further results in a decrease in its transmission performance, as shown in Fig. S3(b). When the degree of rarefaction in the microchannels is kept unchanged, the overall transmission performance increases approximately linearly with the number of parallel microchannels. Specifically, when NPS = 2, 3, 4, and 5, the total transmission performance is $1.33 \times 10^{-7}$ $kg{\cdot}m^{-1}{\cdot}s^{-1}$, $1.97 \times 10^{-7}$ $kg{\cdot}m^{-1}{\cdot}s^{-1}$, $2.55 \times 10^{-7}$ $kg{\cdot}m^{-1}{\cdot}s^{-1}$, and $3.07 \times 10^{-7}$ $kg{\cdot}m^{-1}{\cdot}s^{-1}$, respectively. The transmission performance of a single microchannel is $6.66 \times 10^{-8}$ $kg{\cdot}m^{-1}{\cdot}s^{-1}$, $6.55 \times 10^{-8}$ $kg{\cdot}m^{-1}{\cdot}s^{-1}$, $6.36 \times 10^{-8}$ $kg{\cdot}m^{-1}{\cdot}s^{-1}$, and $6.14 \times 10^{-8}$ $kg{\cdot}m^{-1}{\cdot}s^{-1}$, respectively. These results indicate that, when the microchannel size remains unchanged, the transmission performance of an individual microchannel gradually decreases as the number of parallel microchannels increases.

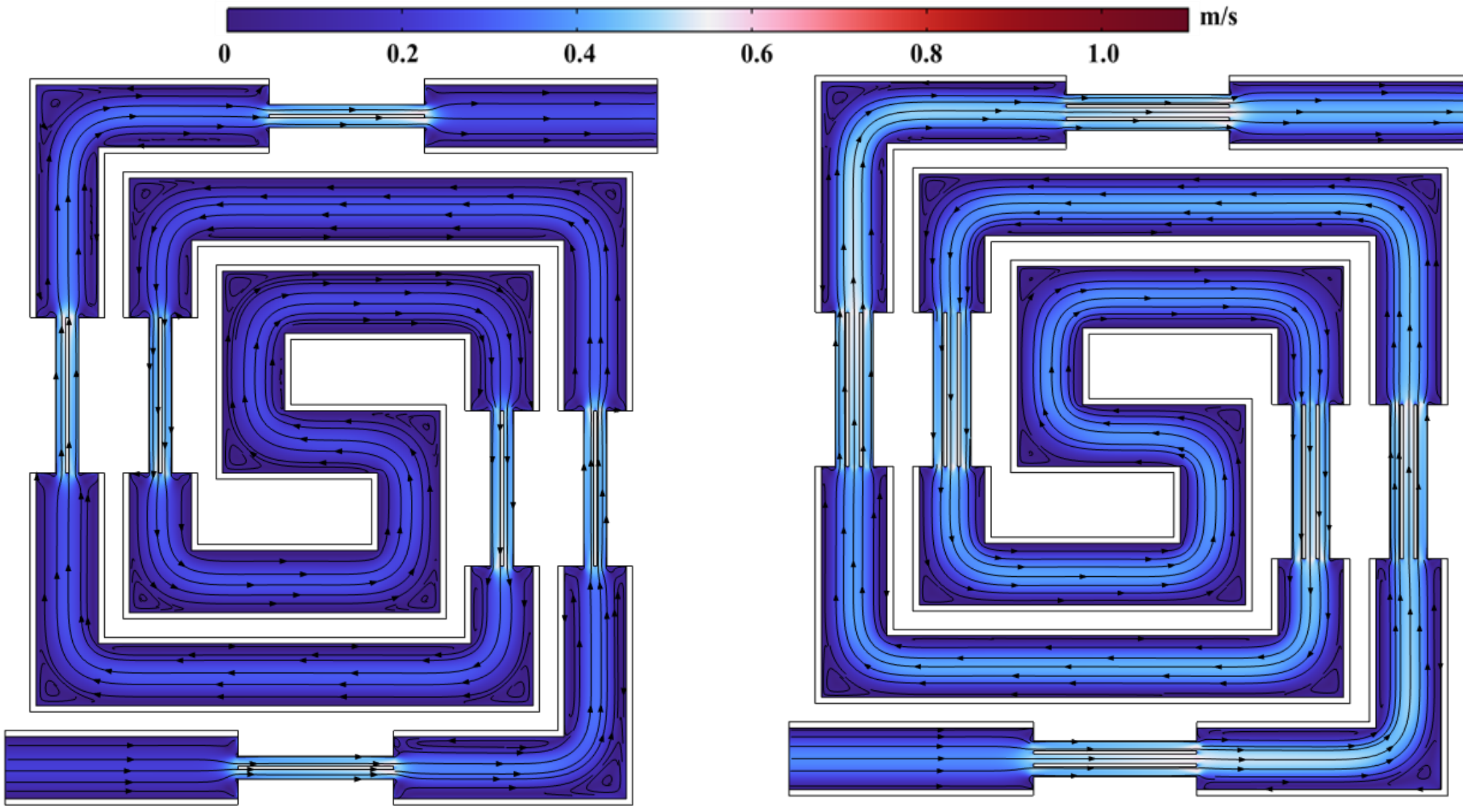

(a) NSS = 6, NPS = 2.

(b) NSS = 6, NPS = 3.

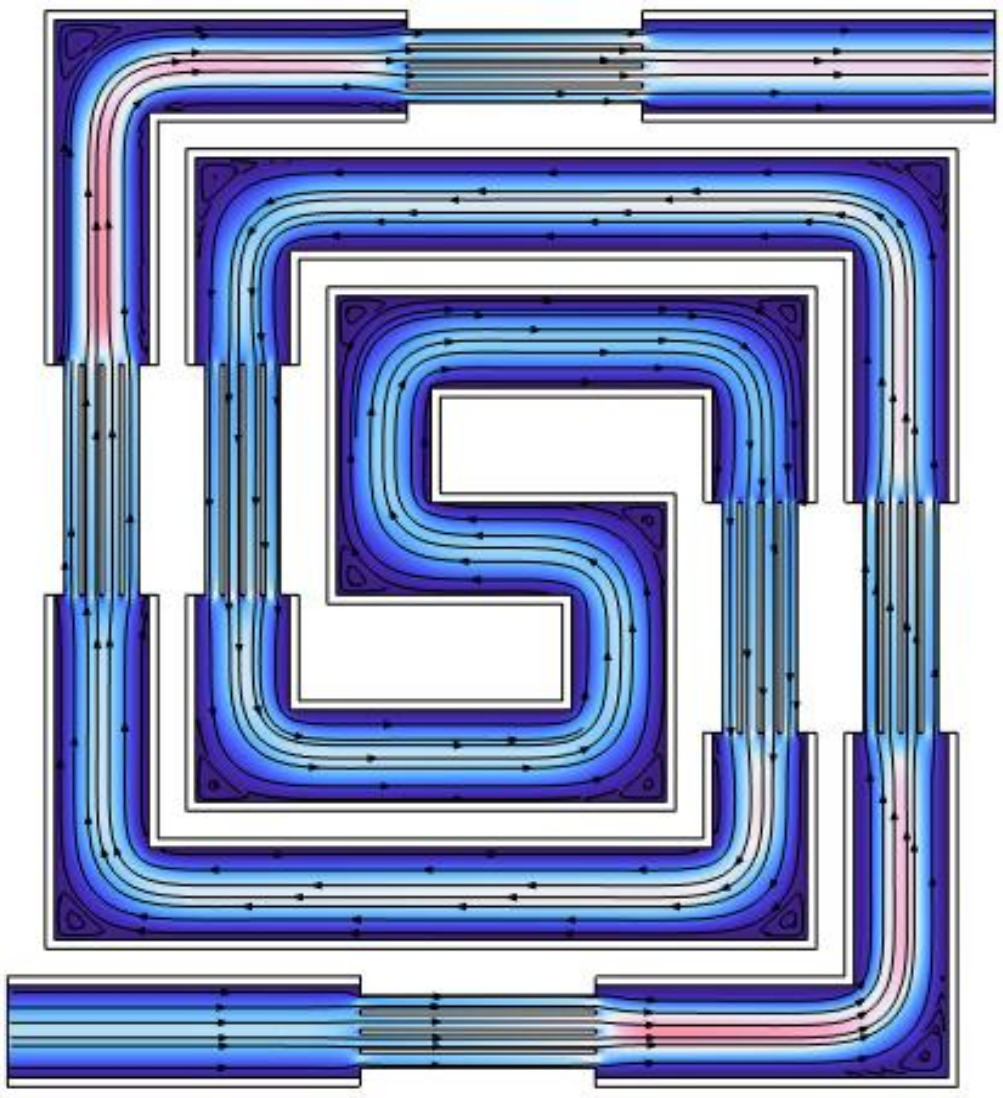

(c) NSS = 6, NPS = 4.

Fig. S2. Contour plot of velocity within the Knudsen pump for different NPS values at a fixed microchannel height of 1.6 μm.

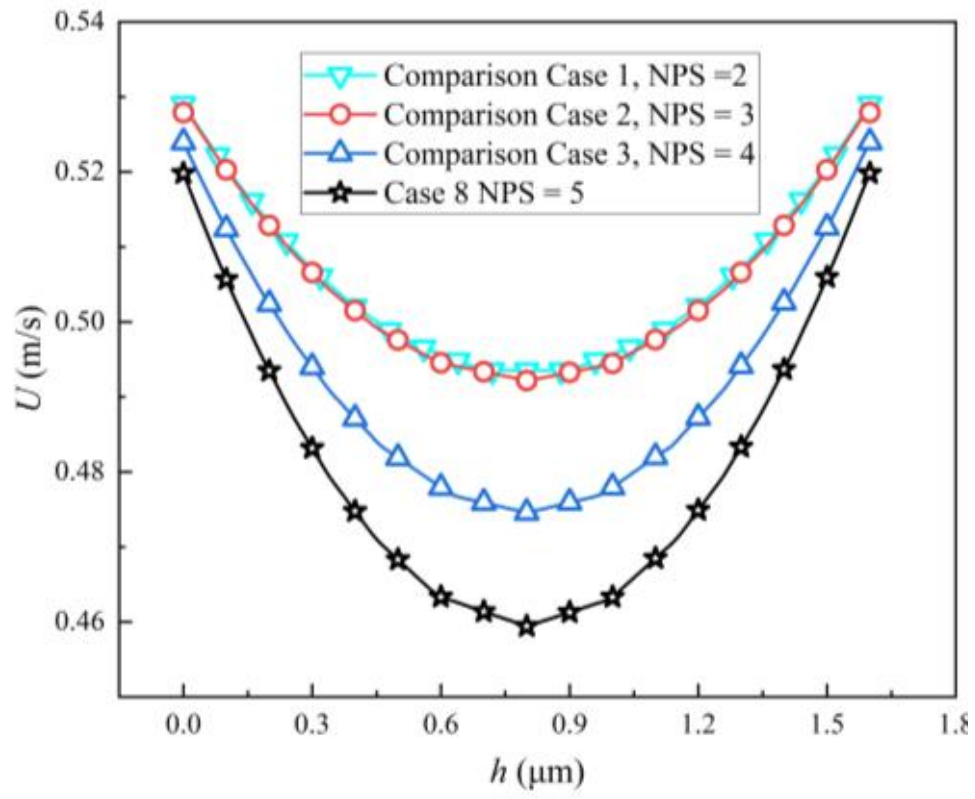

(a) Velocity distributions in the microchannel at Location 1.

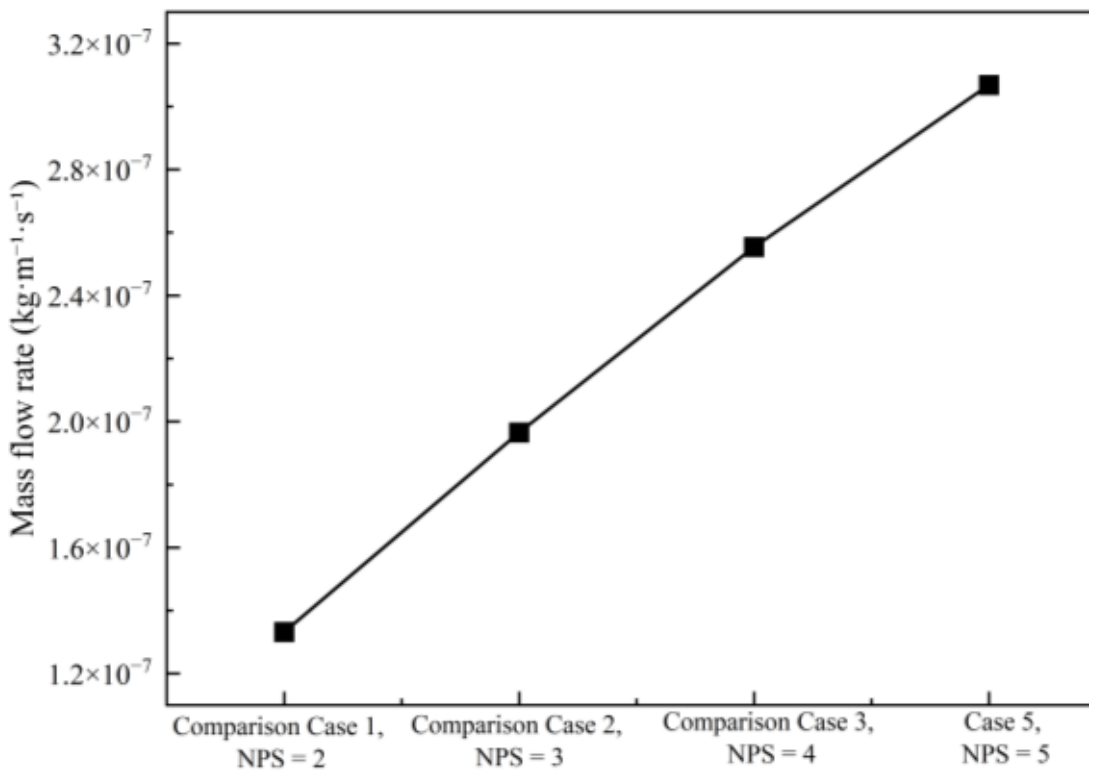

(b) Transmission performance at different NPS.

Fig. S3. Velocity profile in the central microchannel at Location 1 and mass flow rate for different NPS values at a fixed microchannel height of 1.6 μm.

To further analyze the factors affecting the flow and transmission behaviors of the Knudsen pump, Figs. S4 and S5(a) present the temperature distributions within the Knudsen pump for different NPS values. Similar to the temperature distributions observed when the characteristic dimension of the main channel is kept constant, increasing NPS has only a minor effect on the overall temperature distribution within the Knudsen pump when the microchannel characteristic dimension is maintained constant. For the microchannels at each stage, the gas temperature gradually increases from 273 K at the inlet to 323 K at the outlet along the flow direction. In contrast, within the main channel between two adjacent microchannel stages, the gas temperature gradually decreases from 323 K at the outlet of the preceding stage to 273 K at the inlet of the subsequent stage. Therefore, the temperature distributions within both the microchannels and the main channel remain highly consistent across different NPS values. However, unlike the variation in the Knudsen number with NPS observed when the characteristic dimension of the main channel is kept constant, the Knudsen number distributions within the microchannels remain nearly unchanged with increasing NPS when the microchannel characteristic dimension is maintained constant. Along the gas flow direction, the Knudsen number within the microchannels gradually increases from approximately 0.064 at the inlet to approximately 0.078 at the outlet, with essentially the same range of variation for different NPS values. This is primarily because the microchannel characteristic dimension remains constant, while the temperature distributions and absolute pressure levels within the microchannels are also nearly identical across different NPS values. Consequently, the degree of gas rarefaction within the microchannels remains essentially unchanged as NPS increases.

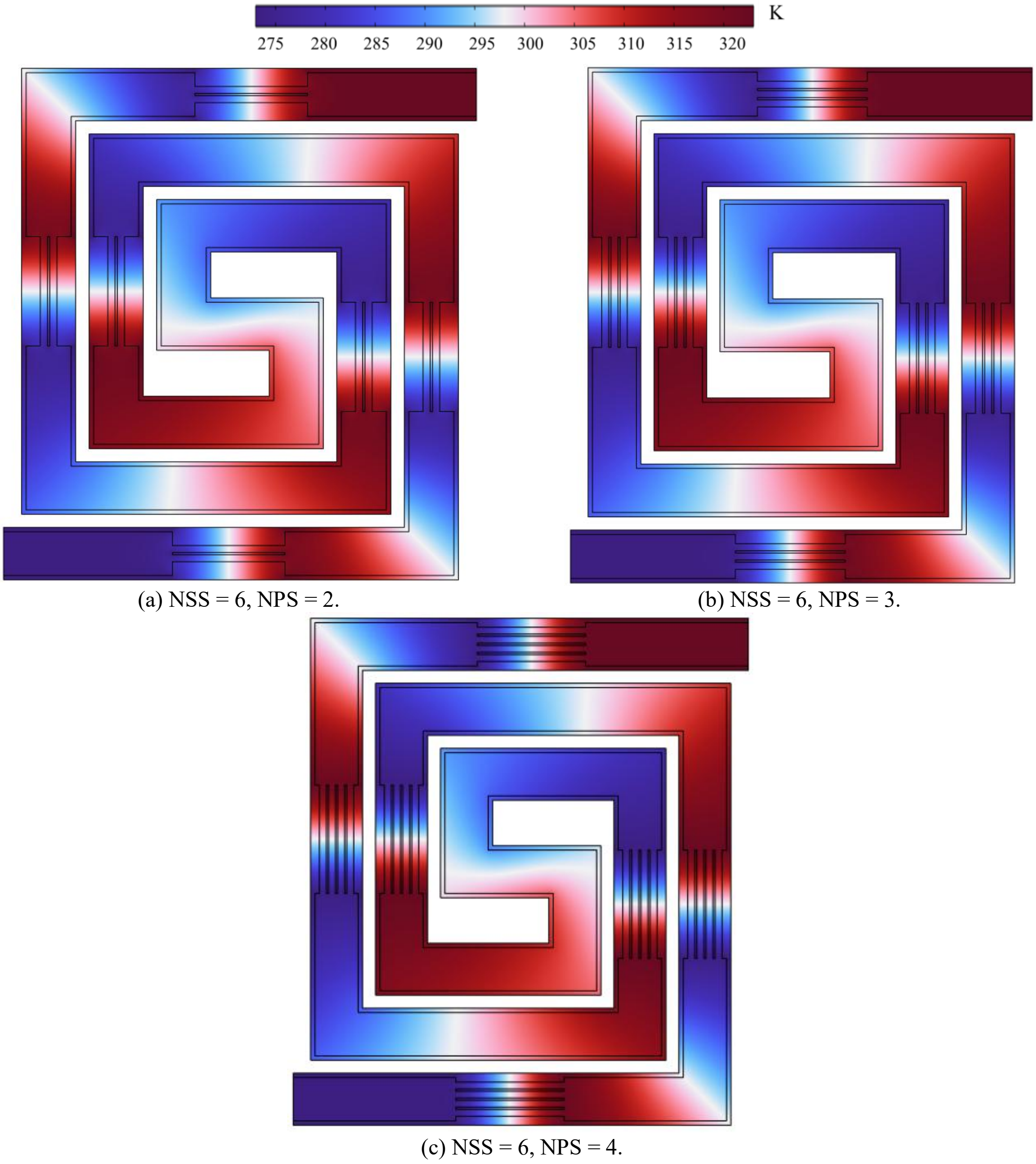

Fig. S4. Temperature contours within the Knudsen pump for different NPS values at a fixed microchannel height of 1.6 μm.

On the other hand, from the perspective of the container-to-microchannel flow cross-sectional area ratio (Fig. S5(b)), the microchannel velocities remain nearly unchanged when the ratio is below 10:4.8. However, when the ratio reaches 10:6.4, the microchannel velocity decreases noticeably. Specifically, at $NPS = 2$ and $NPS = 3$, the average microchannel velocity is approximately 0.528 m/s. When $NPS$ increases to 4 and 5, the average velocity decreases to 0.524 m/s and 0.519 m/s, respectively. Analysis of the pressure distributions shows that, at $NPS = 2$ and $NPS = 3$, the pressure difference across the microchannels remains approximately 23 Pa, with a variation of less than 1 Pa. This indicates that the suppressive effect of the container is relatively weak. When $NPS$ increases to 4 and 5, the pressure difference increases to 29 Pa and 40 Pa, respectively. These results indicate that the pressure loading induced by the main container becomes stronger as the difference between the flow cross-sectional areas of the container and microchannel decreases. The increased pressure loading strengthens the reverse Poiseuille flow opposing thermal transpiration. Consequently, the average microchannel velocity decreases, and the transmission capacity of an individual microchannel is reduced. The flow behavior near the microchannel–container connection provides further insight into this effect. As the flow cross-sectional area of the microchannel becomes comparable to that of the container, the thermal transpiration flow leaving the microchannel becomes increasingly confined by the container walls. The available space for flow adjustment after the gas enters the container is therefore reduced. This promotes local pressure accumulation near the microchannel outlet and increases both the pressure loading across the microchannel and the flow resistance. In contrast, when the flow cross-sectional area of the container is much larger than that of the microchannel, the gas has more space for flow adjustment after entering the container. The influence of the container walls near the microchannel outlet is therefore weaker, resulting in less local pressure accumulation and lower flow resistance. This result is consistent with the findings of Li et al., who reported that containers at both ends of a Knudsen pump induce an additional pressure difference across the microchannels and generate flow resistance opposite to the direction of thermal transpiration, thereby suppressing gas transport through the microchannels [41].

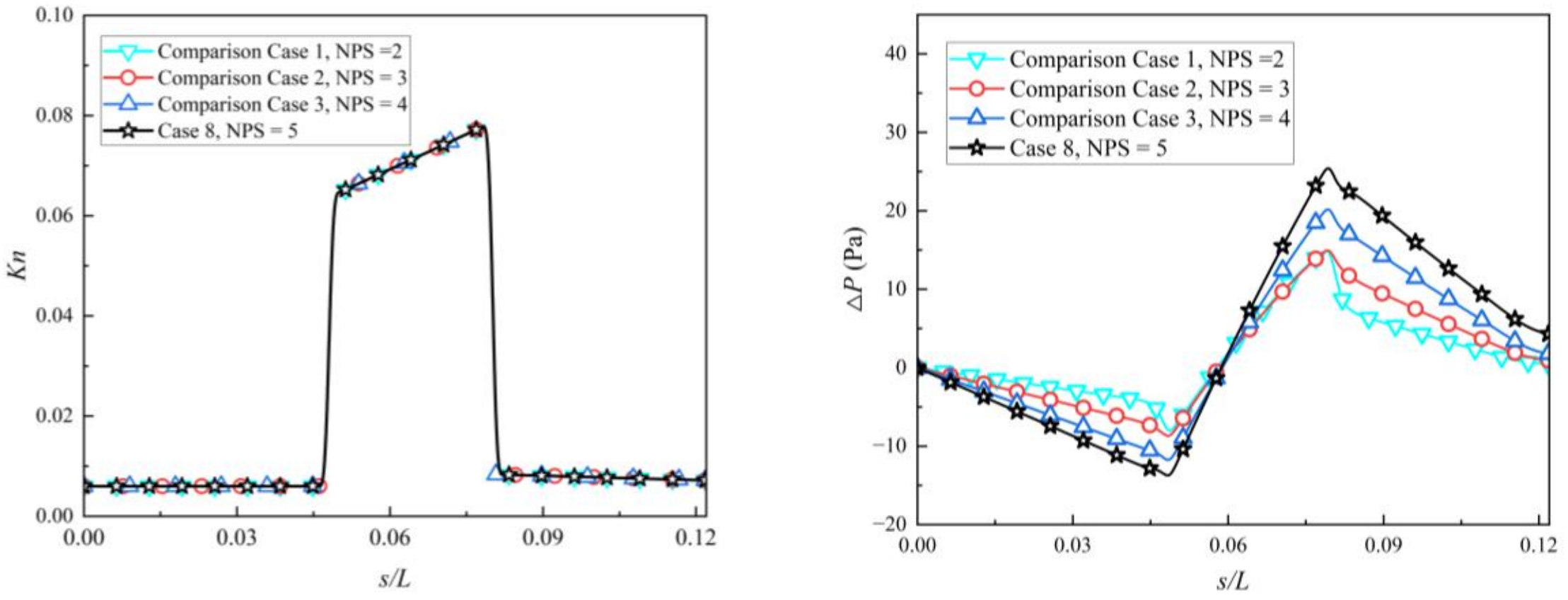

(a) Knudsen number distribution. (b) Pressure distribution.

Fig. S5. Pressure and Knudsen number distributions in the central microchannel at Location 1 for different NPS values at a fixed microchannel height of 1.6 μm.